\documentclass[aps,amsmath,amssymb,prb,twocolumn,superscriptaddress,showpacs,floatfix, longbibliography, reprint]{revtex4-2}
\usepackage{amsmath}
\usepackage{bbold}
\usepackage{braket}
\usepackage{graphicx}
\usepackage{caption}
\usepackage{subcaption}
\usepackage{bm}% bold math
\usepackage{color}
\usepackage{xcolor}
\usepackage[colorlinks=true, urlcolor=blue, linkcolor=red, citecolor=blue]{hyperref}

\newcommand{\taux}{\tau^x}
\newcommand{\tauy}{\tau^y}
\newcommand{\tauz}{\tau^z}
\newcommand{\tauzxxz}{\tau^z\tau^x + \tau^x\tau^z}

\newcommand{\odd}[1]{\textcolor{red}{#1}}
\newcommand{\even}[1]{\textcolor{orange}{#1}}

\begin{document}
\title{Detecting the largest correlations using the correlation density  matrix: a quantum Monte Carlo approach}

\author{Aditya Chincholi} % Please advise on name order here
\affiliation{Univ Toulouse, CNRS, Laboratoire de Physique Th\'eorique, Toulouse,  France.}
\author{Sylvain Capponi} 
\affiliation{Univ Toulouse, CNRS, Laboratoire de Physique Th\'eorique, Toulouse,  France.}
\author{Fabien Alet} 
\affiliation{Univ Toulouse, CNRS, Laboratoire de Physique Th\'eorique, Toulouse,  France.}
\date{\today}

% Abstract needs to out of the twocolumn effect
\begin{abstract}
    We present a quantum Monte Carlo-based approach to detect and compute the most dominant correlations for many-body systems without prior knowledge. It is based on the measurement and analysis of the correlation density matrix between two (small) subsystems embedded in the full (large) sample. In order to benchmark this procedure, we investigate zero-temperature quantum phase transitions in one- and two-dimensional quantum Ising model as well as the two-dimensional bilayer Heisenberg antiferromagnet. The method paves the way for a systematic identification of unknown or exotic order parameters in unexplored phases on large systems accessible to quantum Monte Carlo methods.
\end{abstract}

\maketitle

\section{Introduction}
\label{sec:Introduction}

The automatic detection and identification of different phases of quantum matter is a topic of long-standing interest, even before the advent of machine learning techniques. One outstanding set of questions can be simply formulated as follows: given a quantum state of an interacting system (for instance the ground-state wave-function or the thermal state defined by a density matrix), can one identify if symmetry-breaking takes place? If yes, what is the order parameter ? If not, what are are the leading correlations, and possibly as well are there any traces of non-local order? It is quite crucial to answer these questions in an unbiased way (e.g. for instance without postulating {\it a priori} what are the possible order parameters). This is a very exacting task for the quantum many-body problem as even simple-looking local interactions can give rise to complex, unexpected, order and correlations~\cite{andersonMoreDifferent1972}, in particular when frustration and fluctuations are strong, not to mention the intrinsic exponential difficulty of the many-body problem.

On one hand, machine learning (ML) techniques have growing success (see e.g.~\cite{carrasquilla_machine_2017,vanNieuwenburg_learning_2017,hu_discovering_2017,zhang_machine_2018}) on detecting classical and quantum phases of matter, but they also heavily rely on high-quality input data, training time/quality, and potentially on details of the ML architecture and parameters. The (lack of) interpretability is also sometimes an issue~\cite{krennScientificUnderstandingArtificial2022}. On the other hand, methods based on quantum information insights (such as the scaling of entanglement entropy, mutual information etc) clearly reflect the profound nature of the quantum states and can capture symmetry-broken phases, topological order, phase transitions (for a review, see Ref.~\onlinecite{laflorencieQuantumEntanglementCondensed2016}). However, the central quantity to be used is not always accessible in numerical computations of many-body quantum systems, as for instance the von Neumann entanglement entropy in dimension larger than one. Furthermore, when long-range order is present, these approaches do not necessarily indicate which {\it kind} of order is present, and if not, which correlations dominate the physics. 

A very insightful method was proposed by Henley and coworkers in a series of investigations~\cite{cheongManybodyDensityMatrices2006,cheongCorrelationDensityMatrix2009}. Consider first any two observables ${O}_A$ and ${O}_B$ which have respectively support on two different regions of spaces $A$ and $B$ and their connected correlator in a given quantum state:
\begin{align}
    \braket{ O_A O_B }_c  = \braket{ O_A O_B } - \braket{O_A} \braket{O_B}
\end{align}
where the expectation value is taken in the quantum state. Any of these correlators can be encapsulated by the correlation density matrix (CDM) $\rho^c_{AB}$:
\begin{align}
    \rho^c_{AB} = \rho_{AB} - \rho_A \otimes \rho_B
    \label{eq:CDM}
\end{align}
where the reduced density matrices (RDMs) $\rho_{AB}$, $\rho_A$, $\rho_B$ are obtained by tracing out the degrees of freedom except those on $A \cup B$, $A$ and $B$ respectively. Indeed we simply have:
\begin{align}
    \braket{ O_A O_B }_c  = \mathrm{Tr} \left[ O_A O_B \rho^c_{AB} \right].
\end{align}
The insight by Henley and coworkers is to perform a singular value decomposition (SVD) [See Appendix \ref{appendix:svd}] of the CDM in order to recast it into a sum of \emph{independent}, Frobenius-orthonormal correlations:
\begin{align}
    \rho^c_{AB} = \sum_i \sigma_i X^{(A)}_i \otimes \left( Y^{(B)}_i \right)^\dagger
    \label{eq:cdmdecomp}
\end{align}
where $X^A$, $Y^B$ are operators with support on $A$ and $B$. The operators associated with the largest singular values $\sigma_i$ correspond to the strongest correlations in the system (with the Frobenius norm) -- hence allowing to answer the initial question 'which kind of order is dominant'. We note that similar ideas, but with a different formulation of the methods, can be found in Refs.~\cite{furukawaSystematicDerivationOrder2006,guConstructOrderParameters2013}.

A few in the past have used RDM-based approaches to study the ordering present, primarily through exact diagonalization (ED)~\cite{cheongManybodyDensityMatrices2006,furukawaSystematicDerivationOrder2006}. These studies construct order parameters from the RDMs of symmetry-(un)breaking combinations of degenerate ground states or examine the entanglement spectrum. These were followed by CDM-based studies for separated site clusters using construction of order parameters and examinations of long-range correlations via exact diagonalization~\cite{cheongCorrelationDensityMatrix2009,sudanUncoveringPhysicsFrustrated2010} and density-matrix renormalization group (DMRG) techniques~\cite{munderCorrelationDensityMatrices2010}. Construction of order parameters has also been done via analysis of the mutual information of separated site clusters~\cite{guConstructOrderParameters2013}. These methods, however, are limited in capability when extended to higher dimensions or larger systems sizes. By utilizing quantum Monte Carlo, we seek to bypass this limitation.

\begin{figure}
    \centering
    \includegraphics[width=\linewidth]{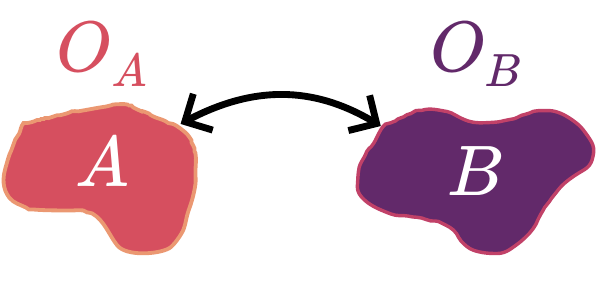}
    \caption{Illustration of correlations between clusters $A$ and $B$ using operators $O_A$ and $O_B$.}
    \label{fig:Gen_ClusterCorrGeneric}
\end{figure}

In this work, we leverage recent developments in stochastic estimates of the reduced density matrix to adapt the CDM to quantum Monte Carlo (QMC). This allows to broaden the scope of the CDM method by bringing the capacity of the QMC method to treat very large systems, in any dimension, and this for a broad range of systems exhibiting a large variety of physical behavior. Being able to tackle large clusters allows to reduce/tame finite-size effects which are often a hurdle for ED, as well as to address critical behavior which is very hard to capture with DMRG. 

We first present in detail how to extract the CDM and the SVD operators $X$ and $Y$ with quantum Monte Carlo, including details on how to concretely estimate error bars (inherent to the stochastic approach) and how to optimally use symmetries to increase the efficiency of the sampling process. The scheme is illustrated through the specific stochastic series expansion~\cite{sandvikStochasticSeriesExpansion1999} (SSE) method, but can easily be adapted to other QMC algorithms. SSE is the state-of-the-art method for interacting quantum spin systems since it provides a robust and reliable approach to sampling quantities of interest in various models~\cite{sandvikStochasticSeriesExpansion2003,wangHighprecisionFinitesizeScaling2006}. Recently, it has been used to calculate the RDM of the ground state of the periodic Heisenberg ladder to obtain the entanglement spectrum~\cite{maoSamplingReducedDensity2025,mao2025detectingemergentcontinuoussymmetry}, and to study negativity and multipartite entanglement in the 2d transverse field Ising model~\cite{wangEntanglementMicroscopyTomography2025}. We will utilize this method to calculate the CDM in our approach.
Numerical results are provided in Sec.~\ref{sec:Results}.
We first apply it to the well-known transverse field Ising model in one dimension, which allows to benchmark our results using available conformal field theory expressions. Then, we consider the same model in two dimensions which has been extensively studied numerically.  Finally we illustrate the method to detect the quantum phase transition on a bilayer Heisenberg model with SU(2) symmetry. Our conclusions and perspectives are given in Sec.~\ref{sec:conclusion}.

\section{Measuring the Correlation Density Matrix with Quantum Monte Carlo methods}
\label{sec:Procedure}

Our strategy to measure the CDM $\rho^c_{AB}$ (Eq.~\ref{eq:CDM}) is by computing matrix elements of $\rho_{AB}$, $\rho_A$ and $\rho_B$ in a given computational basis through Monte Carlo estimates. The size of these matrix grow exponentially with the size of $A$ and $B$. Hence, we will focus on lattice models where $A$ and $B$ are composed of a few sites (typically one to four), such that measuring and storing all independent matrix elements (that are non vanishing or related by symmetries) remains manageable.

We first recall how to measure elements of reduced density matrices in quantum Monte Carlo as recently introduced in Ref.~\cite{maoSamplingReducedDensity2025} in Sec.~\ref{sec:SSE_RDM}. For models exhibiting a higher symmetry (such as SU(2) symmetry), we advance the method by adding an improved estimator. These elements are obtained from Monte Carlo estimates and thus come with error bars. We present in Sec.~\ref{sec:OperatorBasisDecomposition} our strategy for extracting the dominant correlators $X,Y$ (in Eq.~\ref{eq:cdmdecomp}) using a symmetry-based decomposition for operator basis in $A$ and $B$. We also further discuss the use of symmetries to reduce the inherent statistical errors in Sec.~\ref{sec:UseOfSymmetries}.

\subsection{Stochastic Series Expansion for RDMs}
\label{sec:SSE_RDM}

A large set of quantum Monte Carlo (QMC) methods are devised on the sampling of the partition function $Z(\beta)=\mathrm{Tr}\left[ \exp(-\beta H) \right]$ ($\beta=1/k_B T$ being the inverse temperature) on a system described by the Hamiltonian $H$. This is for instance the case of path-integral QMC which performs a sampling in a ($d+1$) dimensional phase space, where the extra-dimension is the imaginary time that goes from $\tau=0$ to $\tau=\beta$. The trace in the definition of $Z$ imposes periodic boundary conditions in this imaginary time direction. Recently, Ref.~\onlinecite{maoSamplingReducedDensity2025} pointed out that by relaxing the periodic boundary conditions constraint in the imaginary-time, one can also have direct access to reduced density matrices (RDM). We illustrate this idea in the following using the SSE formalism, but the discussion will be generic enough such that the method can be implemented in other QMC algorithms.

% This figure may need to be out of the twocolumn effect
\begin{figure*}[]
    \centering
    \begin{subfigure}[b]{0.49\linewidth}
        \centering
        \includegraphics[width=\textwidth]{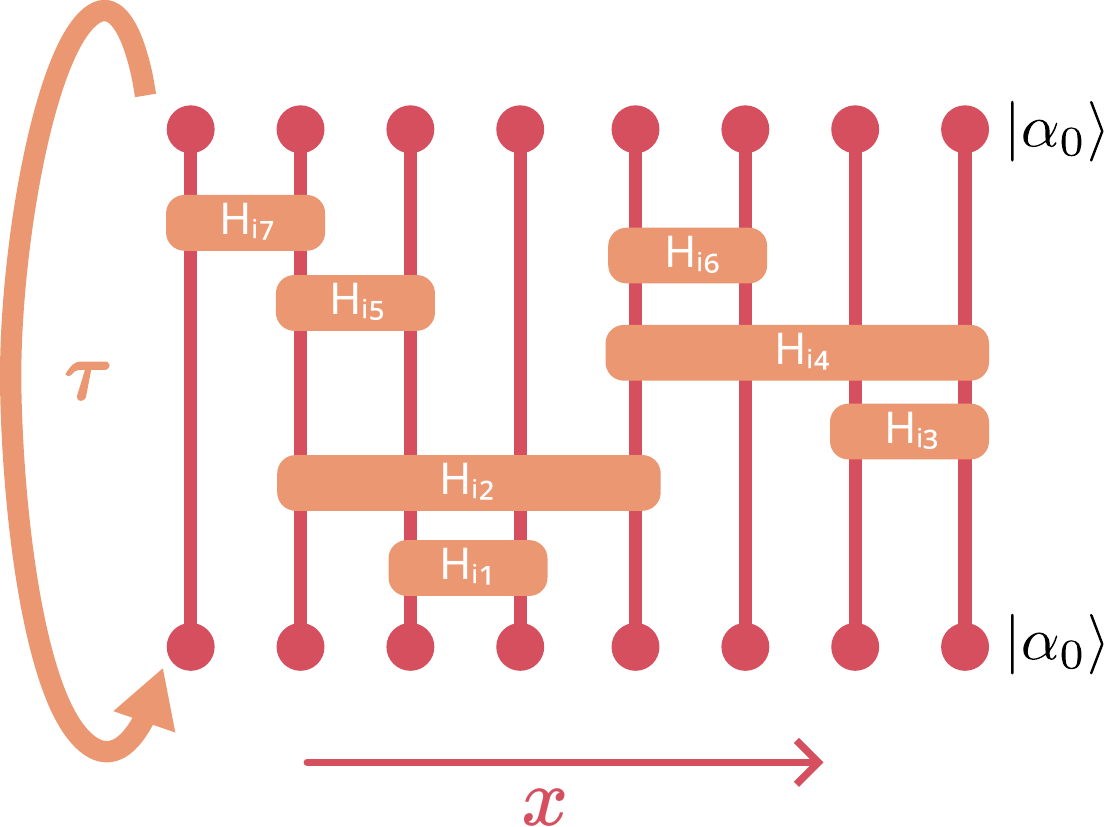}
        \caption{}
        \label{fig:Gen_SSENormal}
    \end{subfigure}
    \hfill
    \begin{subfigure}[b]{0.49\linewidth}
        \centering
        \includegraphics[width=\textwidth]{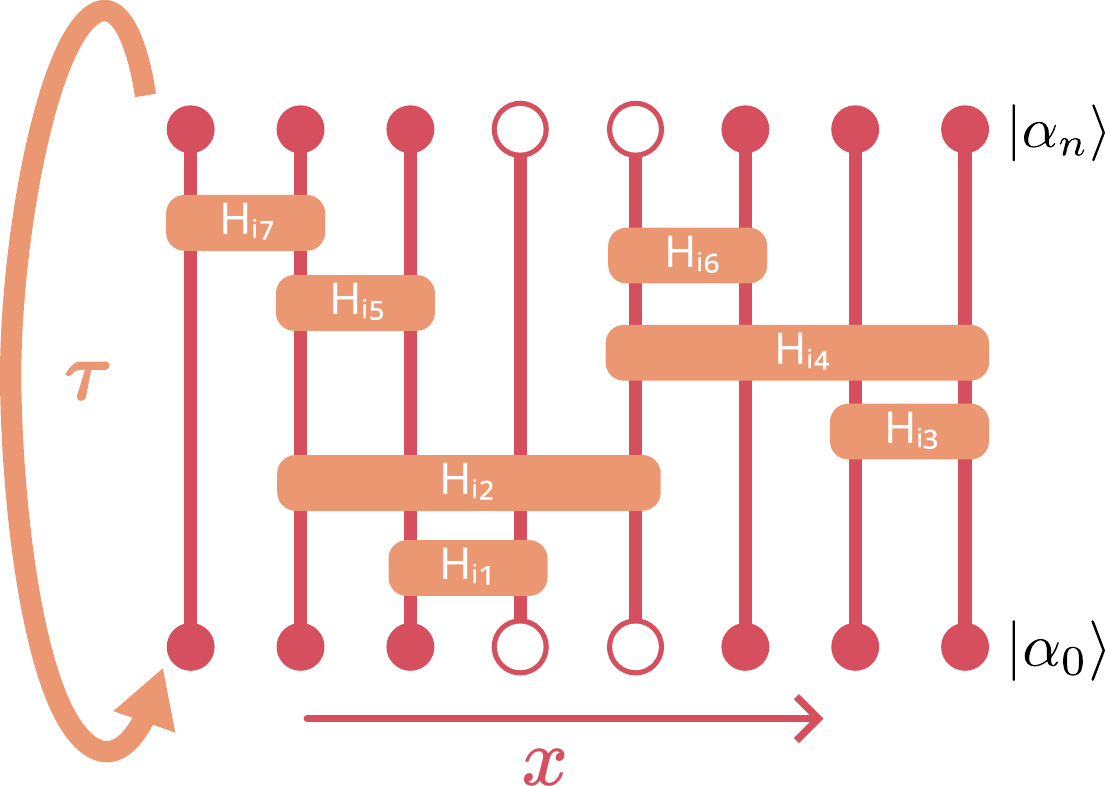}
        \caption{}
        \label{fig:Gen_SSERDM}
    \end{subfigure}
    \caption{Schemes of the partition function sampling \eqref{eq:Z_SSE}: (a) Normal SSE; (b) RDM SSE}
    \label{fig:Gen_SSE}
\end{figure*}

Given a local Hamiltonian $H=-\sum_i H_i$, the SSE method expands the partition function in powers of $\beta$
\begin{align}\label{eq:Z_SSE}
    Z(\beta) &= \sum_{\{\alpha\}} \sum_{S_L}
    \frac{\beta^n (L-n)!}{L!} \nonumber\\
    &\bra{\alpha_0} H_{i_1} \ket{\alpha_1}
    \bra{\alpha_1} H_{i_2} \ket{\alpha_2}
    \bra{\alpha_2}  H_{i_3} \ket{\alpha_3} \\
    &...
    \bra{\alpha_{n-1}} H_{i_n} \ket{\alpha_0}.\nonumber
\end{align}
and samples both the expansion order $n$, the set of operators $H_{i_1}, \dots, H_{i_n}$, and the initial state $\alpha_0$. The expansion has a cutoff at a maximal number (proportional to) $L$. This is illustrated in Fig.~\ref{fig:Gen_SSENormal} for a one-dimensional (1d) system composed of $8$ sites in real space, where the arrow representing periodic boundary conditions in imaginary time indicates the trace condition:  the state $\ket{\alpha_n}$ (after the application of the last local Hamiltonian $H_{i_n}$) is identical to $\ket{\alpha_0}$ (in this figure all filled dots). 

Now consider a (small) subsystem composed of two sites $x_1,x_2$ in the middle of the chain (see Fig.~\ref{fig:Gen_SSERDM}). 
It is easy to convince oneself that by relaxing the periodic boundary conditions (PBC) in imaginary time \emph{only on these two sites} (using open boundary conditions), one has access to the RDM elements
\begin{align}\label{eq:RDM}
    |\rho_{AB}(\beta)_{\tilde{\alpha_0},\tilde{\alpha_n}}| &\propto  \sum_{\{\alpha\}}  \sum_{S_L}
     \frac{\beta^n (L-n)!}{L!} \\
    &\bra{\tilde{\alpha_0}} H_{i_1} \ket{\alpha_1}
    \bra{\alpha_1} H_{i_2} \ket{\alpha_2}
    \bra{\alpha_2}  H_{i_3} \ket{\alpha_3} \nonumber\\
    &...
    \bra{\alpha_{n-1}} H_{i_n} \ket{\tilde{\alpha_n}}.\nonumber
\end{align}
where $\tilde{\alpha_0}, \tilde{\alpha_n}$ are shortcut notations for the system configurations that may differ on $A$ and $B$ sites,  i.e. $\ket{\tilde{\alpha_0}}_{\overline{AB}} = \ket{\tilde{\alpha_n}}_{\overline{AB}}$ but $\ket{\tilde{\alpha_0}}_{AB}$ and $\ket{\tilde{\alpha_n}}_{AB}$ could be different. 

For instance, if the state $\ket{ \tilde{\alpha_0} }_{AB} = \ket{ \uparrow \downarrow }, \ket{ \tilde{\alpha_n} }_{AB} = \ket{ \downarrow \uparrow }$ is observed in a SSE configuration, one counts $+1$ for the measurement of the RDM element $|\rho_{AB}(\beta)_{\uparrow \downarrow ,\downarrow \uparrow}|$. The final estimate is obtained by normalizing by the total number of measurements. Note that the above method allows to measure only the absolute value of RDM matrix elements. The \emph{sign} of RDM elements need to be fixed by symmetry considerations (see a discussion below).

Measuring the full density matrix is impossible in practice, due to the exponential size of the configurations space, but it can be done for the RDM on a not-too-large subsystem.~\footnote{For instance, using translation symmetry, the RDM could be computed up to a 24-site spin-1/2 chain in Ref.~\onlinecite{maoSamplingReducedDensity2025}.}

\subsection{Decomposition of the CDM}

Using the RDM obtained from the SSE procedure, the CDM is constructed via Eq.~\eqref{eq:CDM}. The indices are then rearranged to form the matrix $\tilde{\rho}^c_{(aa'),(bb')}$ where $(xy)$ represents a combined index~\cite{sudanUncoveringPhysicsFrustrated2010,cheongCorrelationDensityMatrix2009}:

\begin{align*}
    \rho^c &= \sum_{a,b,a',b'} \rho^c_{(ab), (a'b')} \ket{a} \ket{b} \bra{a'} \bra{b'} \\
    \tilde{\rho}^c &= \sum_{a,b,a',b'} \rho^c_{(ab), (a'b')} \ket{a} \ket{a'} \bra{b} \bra{b'} \\
    &= \sum_{a,b,a',b'} \tilde{\rho}^c_{(aa'), (bb')} \ket{a} \ket{a'} \bra{b} \bra{b'}
\end{align*}

Upon doing the singular value decomposition (SVD) of $\tilde{\rho}^c_{(aa'),(bb')}$, we obtain 
\begin{align*}
    \tilde{\rho}^c_{(aa'),(bb')} &= U \Sigma V^\dagger \\
    \Sigma &= diag(\sigma_0, \sigma_1, ... \sigma_n) \\
    (X_i^{(A)})_{a,a'} &= U_{(aa'),i} \\
    ((Y_j^{(B)})^\dagger)_{b,b'} &= V^\dagger_{j, (bb')}
\end{align*}

where $\sigma_i$ and $\left\{ X_i^{(A)}, (Y_j^{(B)})^\dagger \right\}$ are the independent connected correlations between $A$ and $B$, and the corresponding operators, as indicated in Eq.~\eqref{eq:cdmdecomp}.

\subsection{Operator basis decomposition}
\label{sec:OperatorBasisDecomposition}

There are two types of information that we obtain from the CDM, namely the singular values $\sigma_i$ and the corresponding operators $X_i$, $Y_i^\dagger$. These operators $X_i$ form a basis for all allowed operators on $A$ (similarly $Y_i^{\dagger}$ on $B$). Crucially, the connected correlation $\braket{X_i Y_i^{\dagger}}_c = \sigma_i$. This means we can automatically obtain the largest scaling operators through this decomposition. We express these operators in another basis $\{P_j\}$, usually chosen to be a physically meaningful basis such as Pauli-strings or SU(2) invariant operators. By using the Frobenius inner product $\mathrm{Tr}\left[ X_i P_j^\dagger\right]$, we can obtain the expansion of these correlations in a basis of our choice. This allows us to determine the most relevant operators in each phase. Such decompositions are also important to group together operators that (to first order) scale in the same manner at critical points.

For example, consider an Ising-like system with operators that can be expressed using strings of the Pauli operators $\{\tau^0=\mathbb{1},\taux,\tauy,\tauz\}$. If $|A|=|B|=1$ (called a 2-site CDM), the operator $X_i$ would be a linear combination of $\taux$, $\tauy$, $\tauz$ and $\mathbb{1}$. For $|A|=|B|=2$ (called a 4-site CDM), $X_0$ would be a linear combination of $\tau^p_1 \tau^q_2$ where $p,q=0,1,2,3$ and subscripts indicate position. For instance, $X_0 = \alpha \tauz_1\taux_2 + \beta \tauy_1 \mathbb{1}_2 + \gamma \tauy_1 \tauz_2$ would be a valid combination. The exact Pauli-strings and coefficients would be dictated by the connected correlations and symmetries present in the system.

\subsection{Use of symmetries}
\label{sec:UseOfSymmetries}

The structure of the RDM and of the CDM are highly influenced by the symmetries present in the system, and these symmetries can be leveraged to improve the statistical estimates of the various quantities.

\emph{Intrinsic symmetries of the Hamiltonian:} The RDM inherits the symmetries of the Hamiltonian. For instance, if there is a spin inversion symmetry $\uparrow \Leftrightarrow \downarrow$, then it can be enforced at the end of the estimations of $\rho^{\mathrm{AB}}$. If the total magnetization is conserved $[ H , S^z ] =0$, the RDM and the CDM have a block-diagonal structure, which can be leveraged to reduce the number of possible operators $X^A,Y^B$ in its SVD decomposition (we provide explicit examples below). This is also true when the system enjoys a higher symmetry such as SU($2$) (we also provide examples for the operator decomposition in that case). Further, in the SU(2) case, we are able to devise an \emph{improved estimator}~\cite{evertzLoopAlgorithm2003} for the measurement of the RDM elements. 

\label{sec:ImprovedEstimator}
Consider for instance the S=1/2 Heisenberg model with SU(2) symmetry: the SSE operator string admits a full-loop decomposition~\cite{evertzLoopAlgorithm2003}, where loops are composed of worldline segments. Loop states can represent with the same probability up or down spins (reflecting the SU(2) symmetry). Loops are, in general, closed and follow periodic boundary conditions in imaginary time, except for the loops touching the states $\ket{\tilde{\alpha_0}}$, $\ket{\tilde{\alpha_n}}$ in the region ${AB}$ which are open (see Fig.~\ref{fig:Gen_SSERDM}). The improved estimator consists in considering loop states rather than spin states, in such a way that we can measure $2^{\ell_{AB}}$ different elements of the RDM for a single loop configuration, where $\ell_{AB}$ is the number of distinct loops that touch the $AB$ region on either $\ket{\alpha_0}$ or $\ket{\alpha_n}$.

Certain symmetries may also constrain the sign of the RDM elements. This is important as the RDM sampling scheme cannot obtain the sign of the elements directly, only the magnitude. In case of the S=1/2 Heisenberg couplings on bipartite lattices, the Marshall sign rule~\cite{marshallAntiferromagnetism1997,LSM} allows us to find the sign of the matrix elements. It states that the sign of the wavefunction coefficient for the ground state of an isotropic antiferromagnet on a connected bipartite lattice in a sector of magnetization $M$ is given by $\ket{\psi} = \sum_m c^{(M)}_m (-1)^{|A|/2 - \sum_i S^z_i} \ket{m}$ (for the S=1/2 case)~\cite{marshallAntiferromagnetism1997} where $\ket{m}$ are the spin configurations in the computational basis. For the Ising models, the Hamiltonian has non-positive off-diagonal matrix (i.e. is stoquastic), therefore, by the Perron-Frobenius theorem, the ground state is positive. As a consequence, all the signs on the RDM are positive.

\emph{Geometric symmetries of the spin clusters:} $A$ and $B$ clusters may be symmetric under exchange or under certain permutations of their sites. In such situations, this type of symmetry can be enforced also at the end of the estimate of $\rho_{AB}$. In our examples shown in the main text, we take $A$ and $B$ to be symmetric under exchange. This leads to the question of how best to estimate $\rho_{AB}$, $\rho_{A}$ and $\rho_{B}$ in such conditions. In our tests, calculating $\rho_{AB}$ and computing $\rho_{A}$ and $\rho_{B}$ through a partial trace yields the best estimates for eliminating any noise and reducing errors. Therefore, this is the procedure that has been followed for all the subsequent examples.  

\emph{Emergent/Unknown symmetries}: These may be recognized from the operator decomposition obtained from the mean CDM. Even though the decomposition may have numerical errors, most symmetries can be seen easily in terms of the deviations being very small provided the corresponding singular value is well-resolved within errors. This signature in the operator decomposition depends on the sensitivity of the symmetry to noise from the QMC procedure and we have not demonstrated any such example in this paper.

While symmetries may be leveraged for improving estimates, they are also crucial in the bootstrap procedure to produce reliable error bars. Since we generate samples from the mean RDM in this procedure to estimate various quantities, the samples generated must also follow the relevant symmetries. This procedure is described in Sec. \ref{sec:Bootstrap}. Without explicit imposition, the bootstrap samples will generally violate symmetries present in the system. The mean CDM will not have these issues. Therefore it is important to identify the requisite symmetries from the mean CDM in order to obtain reliable error bars.

%%%%%%%%%%%%%%%%%%%%%%%%%%%%%%%%%%%%%%%%%%%%%%%%%%%%%%%%%%
\section{Results}
\label{sec:Results}

\subsection{1D Transverse Field Ising Model}
\label{sec:OneDimensionalTransverseFieldIsingModel}
% \onecolumngrid

\begin{figure*}[]
    \centering
    \begin{subfigure}[b]{0.2\linewidth}
        \centering
        \includegraphics[width=\textwidth]{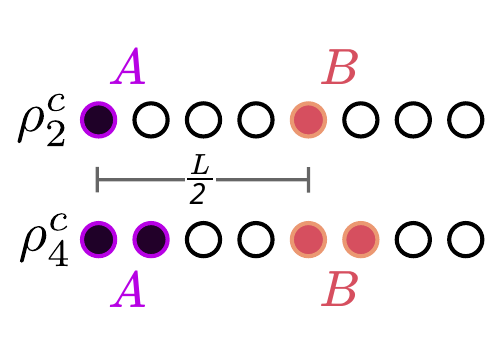}
        \caption{}
        \label{fig:1dTFIM_Layout}
    \end{subfigure}
    \hfill    
    \begin{subfigure}[b]{0.25\linewidth}
        \centering
        \includegraphics[width=\textwidth]{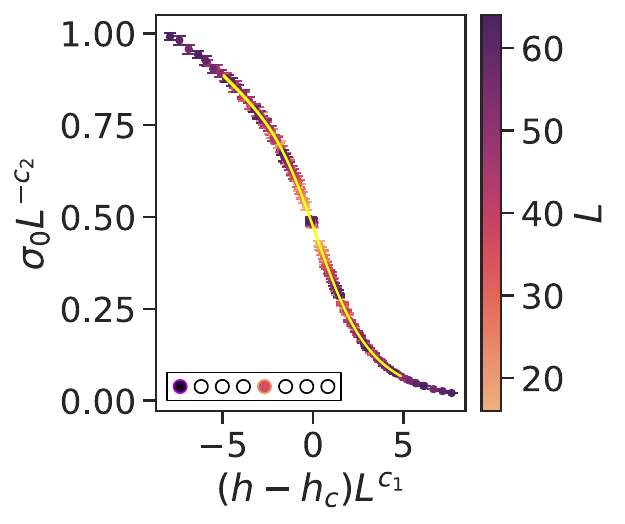}
        \caption{}
        \label{fig:1dTFIM_ScalingHarada}
    \end{subfigure}
%    \vfill
    \begin{subfigure}[b]{0.25\linewidth}
        \centering
        \includegraphics[width=\textwidth]{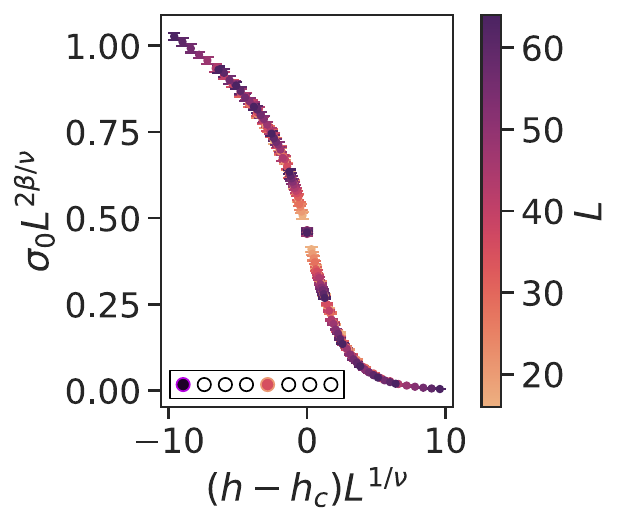}
        \caption{}
        \label{fig:1dTFIM_ScalingKnown}
    \end{subfigure}
    \hfill
    \begin{subfigure}[b]{0.25\linewidth}
        \centering
        \includegraphics[width=\textwidth]{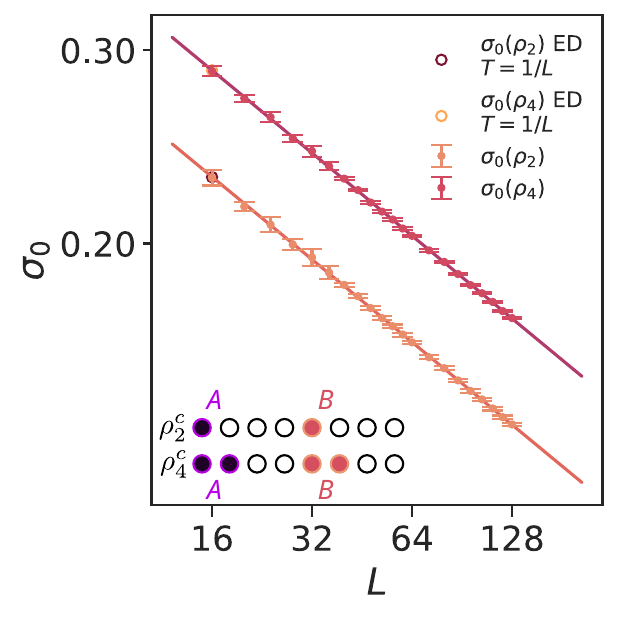}
        \caption{} 
        \label{fig:1dTFIM_SV0vsLength}
    \end{subfigure}
    \caption{1D TFIM: (\ref{fig:1dTFIM_Layout}) Layout of subsystems $A$ and $B$ chosen for the 4-site CDM, (\ref{fig:1dTFIM_ScalingHarada}) Non-linear scaling fit using 2-site CDM yields fitted parameters - $h_c^*=1.00152(97)$ and critical exponents $\nu^* = 0.955(10)$ $\beta^* = 0.1266(49)$~\cite{haradaBayesianInferenceScaling2011}, (\ref{fig:1dTFIM_ScalingKnown}) Scaling collapse for 2-site CDM data using known exact values: $h_c = 1$, $\nu = 1$, $\beta=1/8$ ($\Delta_\sigma = 1/8$, $\Delta_\epsilon = 1$)~\cite{difrancescoConformalFieldTheory1997}, (\ref{fig:1dTFIM_SV0vsLength}) Log-log plot of the dominant singular value $\sigma_0$ from $\rho^c_4$ vs length at the critical point $h_c=1$. Estimated value of critical exponent is $0.2504(8)$. Hollow circles indicate finite temperature ED calculations at $T=1/L$ while filled circles indicate QMC results at the same temperature}
\end{figure*}

% \twocolumngrid

We start by considering the simplest example of quantum phase transitions, namely the one-dimensional Ising model in a transverse field (1D TFIM)~\cite{Dutta_Aeppli_Chakrabarti_Divakaran_Rosenbaum_Sen_2015}:
\begin{equation}
    \label{eq:1DTFIM_Hamiltonian}
    H = -J \sum_{\braket{ij}} \tauz_i \tauz_j - h \sum_i \taux_i
\end{equation}
where $\tau^{\mu}$ are the Pauli operators and $J=1$ is the energy unit. 
This model is well-known to have a quantum phase transition (QPT) at zero-temperature from a trivial paramagnet at large $h$ (all spins being aligned with the field) to a ferromagnet at small $h$, that spontaneously breaks the $\mathbb{Z}_2$ symmetry corresponding to spin inversion. The exact solution leads to $h_c=1$ and the QPT is in the $1+1$ Ising universality class.~\cite{Dutta_Aeppli_Chakrabarti_Divakaran_Rosenbaum_Sen_2015}

Suppose that we have no prior knowledge about the magnetic order (or its absence) in the ground state of $H$. Through this example, we will demonstrate the power of the QMC-enabled CDM approach to understand the rich physics involved in the phase diagram and at the quantum critical point. We will first discuss how a general look at the variation of singular values can give us a first estimate of the phase diagram, then continue with a more precise characterization of the phase with the dominant operators. We also discuss the critical scalings obtained at the quantum critical point. Finally we comment on subtle details on how to determine numerically the dominant operators and singular values within the QMC scheme that we propose.

In the following, we work in the $\tauz$ computational basis to obtain the matrix elements of the CDM, which is computed for two different geometries represented in Fig.~\ref{fig:1dTFIM_Layout}: (i) $A$ and $B$ being both single sites, separated by $L/2$ (the maximal distance for a chain with periodic boundary conditions) allowing us to obtain $\rho^c_2$, and (ii) $A$ and $B$ being both two consecutive sites, also separated by $L/2$, to access $\rho^c_4$.  

%The results from $\rho^c_2$ are often easier to interpret whereas $\rho^c_4$ yields harder to interpret but richer insights.

\begin{figure}[]
    \centering
    \begin{subfigure}[b]{0.49\linewidth}
%        \centering
        \includegraphics[width=\textwidth]{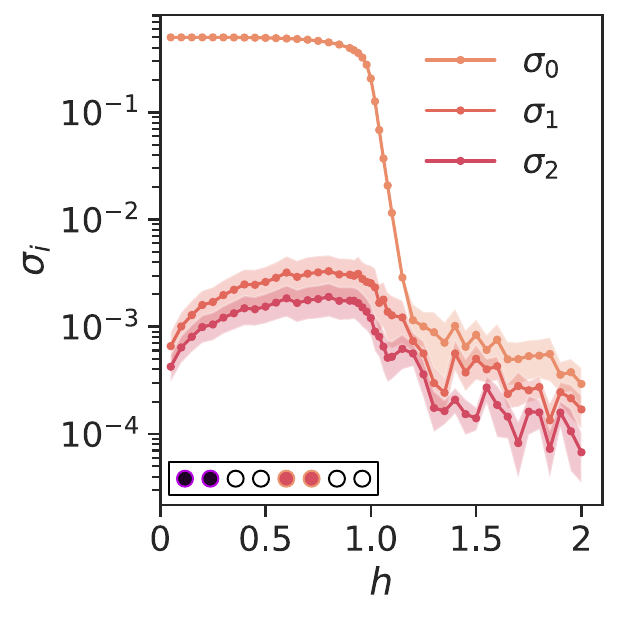}
        \label{fig:1DTFIM_SVvsField_4Site}
        \caption{}
    \end{subfigure}
    \hfill
    \begin{subfigure}[b]{0.49\linewidth}
%        \centering
        \includegraphics[width=\textwidth]{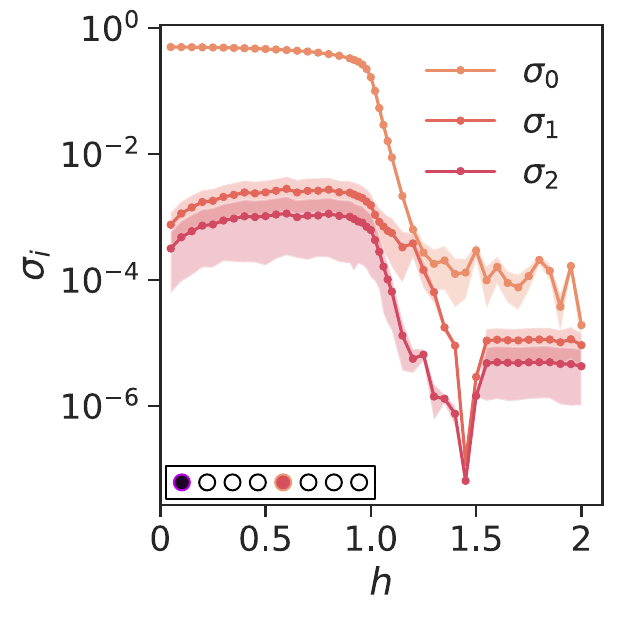}
        \label{fig:1DTFIM_SVvsField_2Site}
        \caption{}
    \end{subfigure}
    \caption{1D TFIM: Singular values vs field ($L=60$) (a) $\rho_4^c$ (smaller singular values are not shown) and (b) $\rho_2^c$. The height of the shaded portion corresponds to the error in the singular values.}
    \label{fig:1dTFIM_SVvsField}
\end{figure}

\emph{Singular values --- } We start the analysis by considering the first few dominant singular values in the SVD of $\rho^c$, which are represented as a function of the field $h$ in Fig.~\ref{fig:1dTFIM_SVvsField} for a \emph{single} system size $L=60$. These singular values correspond to the largest connected long-distance correlations in the system. For each value of $h$ and by convention, we denote $\sigma_0$ the largest singular value, $\sigma_1$ the next largest one etc. Note however that it is possible (and this will indeed be observed) that the physical operator associated to each $\sigma_i$ changes with the field, corresponding to effective level crossings in the SVD spectrum as a function of $h$. 

The behavior of $\rho_4^c$ (left panel) and $\rho_2^c$ (right panel) is very similar: at low values of the field $h$, the CDM is dominated by a single singular value (with $\sigma_0$ of $O(1)$) while lower singular values are orders of magnitude smaller. Near $h \sim 1$, an abrupt change takes place, with $\sigma_0$ decreasing rapidly as $h$ increases until a level crossing. As will be explained in details later, this change of behavior in the singular values reflects the quantum phase transition at $h_c=1$ (due to the finite sample size $L=60$, the drop down of $\sigma_0$ does not occur exactly at $h_c=1$).  

\emph{Dominant operators --- } 
Before focusing on the critical point, we first characterize the two phases by analyzing the operators $X^{(A)}_i$ (and $(Y^{(B)}_j)^\dagger$) obtained from the SVD. Since our system is symmetric under exchange of $A$ and $B$, we anticipate that $X^{(A)}_i$ and $(Y^{(B)}_j)^\dagger$ differ at most by an overall sign. Hence, it suffices to consider just $X^{(A)}_i$. The only further remaining symmetry in $\rho_{AB}$ is the global $\mathbb{Z}_2$ symmetry (parity under the operator $\prod_r \taux_r$), which is readily tracked by using the basis formed by the Pauli operator strings ($\{\tau^\mu\}$ for $\rho^c_2$ and $\{\tau^\mu \tau^\nu\}$ for $\rho^c_4$). Each $X^{(A)}_i$ is expanded in this operator basis using the Frobenius scalar product $\mathrm{Tr}[X^{(A)}_i P^\dagger]$ where $P$ is the Pauli operator string.

\begin{table}[]
    \begin{tabular}{|c|p{0.12\linewidth}p{0.18\linewidth}|p{0.12\linewidth}p{0.18\linewidth}|p{0.12\linewidth}p{0.18\linewidth}|}
        \hline
         &
          \multicolumn{2}{c|}{\textbf{Ordered} $\mathbf{h = 0.3}$} &
          \multicolumn{2}{c|}{\textbf{Critical} $\mathbf{h = 1}$} &
          \multicolumn{2}{c|}{\textbf{Disordered} $\mathbf{h = 2}$} \\ \hline
        $X_0$ &
          \multicolumn{1}{p{0.12\linewidth}|}{$1.0^{\#} \odd{ \frac{\tauz}{\sqrt{2}} }$ } &
          $0.4882(23)$ &
          \multicolumn{1}{p{0.12\linewidth}|}{$1.0^\# \odd{ \frac{\tauz}{\sqrt{2}} }$ } &
          $0.08727(18)$ &
          \multicolumn{1}{p{0.12\linewidth}|}{$1.0^\# \even{ \frac{\taux}{\sqrt{2}} }$ } &
          $1.5(2) \times 10^{-4}$ \\ \hline
        $X_1$ &
          \multicolumn{1}{p{0.12\linewidth}|}{$1.0^* \even{ \frac{\taux}{\sqrt{2}} }$ } &
          $0.0017(9)$ &
          \multicolumn{1}{p{0.12\linewidth}|}{$1.0i^* \odd{ \frac{\tauy}{\sqrt{2}} }$ } &
          $0.0013(7)$ &
          \multicolumn{1}{p{0.12\linewidth}|}{$1.0^* \odd{ \frac{\tauz}{\sqrt{2}} }$ } &
          $1.1(6) \times 10^{-5}$ \\ \hline
        $X_2$ &
          \multicolumn{1}{p{0.12\linewidth}|}{$1.0i^* \odd{ \frac{\tauy}{\sqrt{2}} }$ } &
          $7(6) \times 10^{-4}$ &
          \multicolumn{1}{p{0.12\linewidth}|}{$1.0^* \even{ \frac{\taux}{\sqrt{2}} }$ } &
          $5(4) \times 10^{-4}$ &
          \multicolumn{1}{p{0.12\linewidth}|}{$1.0i^* \even{ \frac{\tauy}{\sqrt{2}} }$} &
          $5(4) \times 10^{-6}$ \\ \hline   
    \end{tabular}
    \caption{1D TFIM: Operator decomposition and corresponding singular value from $\rho^c_2$ SVD for $L=48$. $*$ represents that the error bar is of the order of the mean value, $\#$ indicates that there is no error bar within our bootstrap sampling procedure. Orange/Red colors indicate respectively even/odd operators with respect to the global $\mathbb{Z}_2$ symmetry.}
    \label{tab:1dTFIM_OpDecomp_2Site}
\end{table}

\begin{figure}[]
    \centering
    \includegraphics[width=\linewidth]{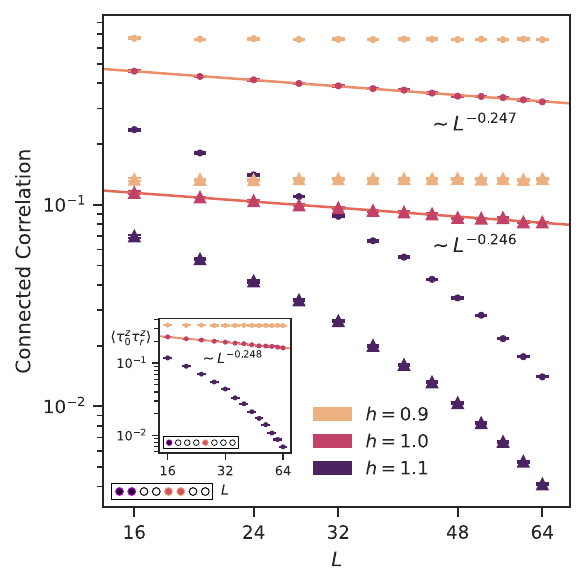}
    \caption{Ising 1D: Physical correlations vs length for different field values for $\rho_4^c$ and $\rho_2^c$ in the 1D TFIM. The circles represent $\frac{1}{4}\braket{(\tau^z_0\mathbb{1}_1+\mathbb{1}_0\tau^z_1)(\tau^z_{L/2}\mathbb{1}_{L/2+1}+\mathbb{1}_{L/2}\tau^z_{L/2+1})}_c$ and triangles represent $\frac{1}{4}\braket{(\tau^z_0\tau^x_1+\tau^x_0\tau^z_1)(\tau^z_{L/2}\tau^x_{L/2+1}+\tau^x_{L/2}\tau^z_{L/2+1})}_c$} 
    \label{fig:1dTFIM_PhysCorrs}
\end{figure}

For $\rho_2^c$, the decomposition (see Table~\ref{tab:1dTFIM_OpDecomp_2Site}) gives us a straightforward interpretation in terms of the $C^{zz}(L)=\braket{\tauz(0)\tauz(L/2)}_c$, i.e., the connected correlation of the magnetization order parameter, corresponding to the well-known $\mathbb{Z}_2$ symmetry breaking in the ordered phase. A finite-size scaling analysis indicates that, in the thermodynamic limit, it is constant in the ordered phase (equal to the square of the order parameter) while it vanishes in the disordered phase alongside all the other correlations. For $\rho_4^c$, the decomposition (see Table~\ref{tab:1dTFIM_OpDecomp_4Site}) reveals a more intricate structure: the dominant correlation is a linear combination of the $\tauz$ and $(\taux\tauz + \tauz\taux)$ operators which are indeed expected to behave similarly since: ($C^{(z+z)(z+z)}(r) = \braket{(\tau^z_0\mathbb{1}_1+\mathbb{1}_0\tau^z_1)(\tau^z_{L/2}\mathbb{1}_{L/2+1}+\mathbb{1}_{L/2}\tau^z_{L/2+1})}_c \sim 4 C^{zz}(r)$ for large $r$).
This suggests that $\tauz$ and $\tauzxxz$ correlations scale similarly. This is verified by calculating $C^{zz}(r) = \braket{\tauz_0\tauz_r}_c$ and $C^{(zx+xz)(zx+xz)}(r) = \braket{(\tauz_0\taux_1+\taux_0\tauz_1)(\tauz_r\taux_{r+1}+\taux_r\tauz_{r+1})}_c$ from the CDM using $\mathrm{Tr}[\rho^c O_0 O_r]$ as shown in Fig.~\ref{fig:1dTFIM_PhysCorrs}.

% \onecolumngrid

\begin{table*}[]
    \centering
    \begin{tabular}{|c|p{0.2\textwidth}p{0.10\textwidth}|p{0.2\textwidth}p{0.10\textwidth}|p{0.2\textwidth}p{0.10\textwidth}|}
        \hline
        \textbf{Op} &
          \multicolumn{2}{l|}{\textbf{Ordered} $\mathbf{h = 0.3}$} &
          \multicolumn{2}{l|}{\textbf{Critical} $\mathbf{h = 1}$} &
          \multicolumn{2}{l|}{\textbf{Disordered} $\mathbf{h = 2}$} \\ \hline
        $X_{0}$ &
          \multicolumn{1}{p{0.22\textwidth}|}{$0.69926(5) \odd{(\mathbb{1}\tauz + \tauz\mathbb{1})/2} + 0.1050(3) \odd{(\taux\tauz + \tauz\taux)/2}$} &
          % 0.4993(12) &
          0.499(1) &
          \multicolumn{1}{p{0.22\textwidth}|}{$0.6326(2) \odd{(\tauz\mathbb{1} + \mathbb{1}\tauz)/2} + 0.3160(4) \odd{(\tauz\taux + \taux\tauz)/2}$} &
          % 0.2157(15) &
          0.216(2) &
          \multicolumn{1}{p{0.22\textwidth}|}{$-0.461^* \even{\taux\taux/2} - 0.554^* \even{\tauz\tauz/2} + 0.493^* \even{\tauy\tauy/2} + 0.344^* (\even{\taux\mathbb{1} + \mathbb{1}\taux)/2}$} &
          0.00034(9) \\ \hline
        $X_{1}$ &
          \multicolumn{1}{p{0.22\textwidth}|}{$-0.635^* \even{(\taux\mathbb{1} + \mathbb{1}\taux)/2} + 0.318^* \even{\taux\taux/2} - 0.204^* \even{\tauy\tauy/2} - 0.14^* \even{\tauz\tauz/2}$} &
          0.0017(6) &
          \multicolumn{1}{p{0.22\textwidth}|}{$-0.410^* \even{(\taux\mathbb{1} + \mathbb{1}\taux)/2} - 0.647^* \even{\taux\taux/2} + 0.273^* \even{\tauy\tauy/2} + 0.413^* \even{\tauz\tauz/2}$} &
          % 0.0023(10) &
          0.002(1) &
          \multicolumn{1}{p{0.22\textwidth}|}{$0.707i^* \even{(\tauy\tauz + \tauz\tauy)/2}$} &
          0.00025(4) \\ \hline
        $X_{2}$ &
          \multicolumn{1}{p{0.22\textwidth}|}{$0.707^* \odd{(\taux\mathbb{1} - \mathbb{1}\taux)/2}$} &
          % 0.00101(29) &
          0.0010(3) &
          \multicolumn{1}{p{0.22\textwidth}|}{$0.386i^* \odd{(\tauy\mathbb{1} + \mathbb{1}\tauy)/2} + 0.592i^* \odd{(\tauy\taux + \taux\tauy)/2}$} &
          0.0011(4) &
          \multicolumn{1}{p{0.22\textwidth}|}{$0.490^* \even{(\mathbb{1}\taux + \taux\mathbb{1})/2} + 0.143^* \even{\taux\taux/2} + 0.026^* \even{\tauy\tauy/2} - 0.706^* \even{\tauz\tauz/2}$} &
          0.00013(5) \\ \hline
    \end{tabular}
    
    \caption{1D TFIM: Operator decomposition and corresponding singular from $\rho^c_4$ SVD in the ordered phase, critical point and disordered phase for $L=48$. Conventions are identical to Table~\ref{tab:1dTFIM_OpDecomp_2Site}. }
    \label{tab:1dTFIM_OpDecomp_4Site}
\end{table*}
% \twocolumngrid

In general, the distinction between phases and characterization of the phases as ``ordered" or ``disordered" requires looking at how the singular values scale with system size. A phase with long-range ordering would show the dominant correlations converging to a constant value whereas the dominant correlations in a phase without long-range order would decrease with increasing system size. This is most easily observed by plotting $\sigma_0(L)$ for different parameter values. The phase transition and the nature of the phases can then be observed by the asymptotic behaviour and the formation of a separatrix, see Fig.~\ref{fig:1dTFIM_PhysCorrs}. Moreover, at the transition, the dominant correlations decay as a power-law with an exponent in excellent agreement with the known value $2\beta=1/4$.~\footnote{The critical exponent related to the order parameter is commonly denoted as $\beta$, not to be confused with the inverse temperature. The distinction should be clear from the context.}

Near the critical point, correlator scaling is governed by the conformal field theory (CFT) present at the critical point.~\cite{difrancescoConformalFieldTheory1997} Correlations scale with system size according to power laws. The scaling ansatz employed is $Q = L^{c_2} f((g - g_c) L^{c_1})$ where $Q$ denotes the observed correlation, $L$ the system length, and $g$ the coupling strength driving the phase transition~\cite{haradaBayesianInferenceScaling2011}. The fitting parameters $c_1$, $c_2$, and $g_c$ represent combinations of critical exponents and the critical coupling strength, respectively. Since all relevant critical exponents can be expressed in terms of two independent exponents, fitting $c_1$ and $c_2$ suffices. 

$\frac{1}{c_1}$ is compared against the scaling exponent of the correlation length $\nu$ and $\frac{c_2}{2}$ is compared against the known values of the operator scaling dimension $\Delta$. In the case of the correlation involved in the phase transition, $2\Delta = \frac{2\beta}{\nu} = d + z - 2 + \eta$ via the scaling and hyperscaling relations, where $\eta$ is the critical exponent encapsulating the decay of the connected correlation of the order parameter at critical point with distance~\cite{kardarStatisticalPhysicsFields2007a,sachdevQuantumPhaseTransitions2011}. For the 1D TFIM $d + z=2$ while $d + z=3$ for the 2D TFIM and the Heisenberg bilayer, since the dynamical exponent $z=1$. For the 1D TFIM, the CFT predicts that the correlator of the order parameter scales as $\displaystyle 2\Delta_{\sigma} = \frac{2\beta}{\nu} = \frac{1}{4}$ at the critical point and the correlation length scales with $\nu = 1$~\cite{difrancescoConformalFieldTheory1997}. 

With a clearer understanding of the phases involved, we shift focus to the critical point. To locate the critical point, we analyze $\rho^c$ as a function of system size. Specifically, we plot the leading singular value $\sigma_0$ of $\rho^c_2$ as a function of system size and field strength. Using a standard scaling analysis mentioned above we achieve an excellent data collapse, see Fig.~\ref{fig:1dTFIM_ScalingHarada}, in full agreement with the exact values of $h_c=1$ and $\nu=1$.

Consequently, we can estimate the ratio of some critical exponents (namely $2\beta/\nu$) from the scaling of $\sigma_0$ at the critical point, avoiding complications associated with non-linear curve fitting. This approach yields a robust estimate consistent with the exact known value of $2\beta/\nu = 1/4$, see Fig.~\ref{fig:1dTFIM_SV0vsLength}.

We could also analyze the smaller singular values $\{\sigma_1, \sigma_2 \ldots\}$. However, in this model the correlations below the dominant one are very small and prone to numerical errors and artefacts. In particular, we see that $\{\sigma_{1+}\}$ have overlapping error bars. This situation can arise due to two causes:

\emph{Degeneracies ---} Some underlying symmetry (known or unknown) can cause different singular values to be equal. We conjecture that this can be a reasonable test for symmetries arising in the field theory by looking at the mean CDM. The error bars will generally be inaccurate if the symmetry is not recognized and imposed during the sampling procedure for the bootstrap. Therefore, one must look for such degeneracies in well-defined singular values (defined below) in the mean CDM.

\emph{Ill-defined Values ---} The QMC sampling process inherently begets an error bar for each matrix element sampled. As a result, the smaller the singular value, the greater the ratio of the error to the mean value. Beyond some value of $n$, all the $\sigma_n$ will suffer from the error bar being of the same order as the mean value. This is the case for $\sigma_{1+}$ in the 1d Ising model. We will refer to such values as ``ill-defined" or ``not well-defined" (and ``well-defined" for the vice versa). In the operator decomposition tables, the corresponding decomposition coefficients are referenced by ``*". As we will see below, the decomposition of these operators can still tell us a bit about the symmetries and structure of the operators in the CFT. 

We now turn our attention to the operator decomposition of $\{X^{(A)}_i\}$. The decompositions at $h=h_c$, given in Table \ref{tab:1dTFIM_OpDecomp_4Site}, reveal the symmetry and scalings of the constituent operators. These decompositions are linear combinations of our basis (Pauli operator strings in this case). A generic linear combination such as $\kappa K + \gamma G$ along with an associated singular value $\sigma$ can imply multiple things depending on the context:

If the value of $\sigma$ is well-defined, then $K$ and $G$ are good indicators of the phase. If this is at the critical point, then they represent the scaling of a CFT operator. Moreover, both $K$ and $G$ must scale with the same CFT operator to leading order. This can be seen by the following argument - if the linear combination of $K$ and $G$ is obtained for all system sizes, then the correlators $\braket{K(0)K(r)}_c$ and $\braket{G(0)G(r)}_c$ must scale as $L^{\alpha}$ where $\alpha$ is twice the scaling dimension of their leading CFT operator. Otherwise, the combination would not persist for increasing $L$. Consider as an example the decomposition of $X^{(A)}_0$ in terms of $\tauz$ and $\tauzxxz$. Given that this decomposition remains stable with increasing system size, this implies that the leading order scaling of both $\tauz$ and $\tauzxxz$ is the same. By comparing with the known expansions of the lattice operators in terms of CFT operators (given in Table \ref{tab:1DTFIM_OPE}), we see that they indeed both scale with the dominant $\sigma_{CFT}$ primary operator to the leading order.

If the value of $\sigma$ is ill-defined, then the scaling of the correlators of $K$ and $G$ loses meaning since we cannot quantify it. However, we can still utilize symmetry arguments irrespective of $\sigma$ being well-defined or not. Owing to the fact that the SVD procedure preserves the block structure of the matrix, $K$ and $G$ must be from the same symmetry sector. We observe that despite $\sigma$ being ill-defined, $K$ and $G$ share the symmetry properties of their associated CFT operator. For example, $X_1$ is formed from $\taux$, $\taux\taux$, $\tauy\tauy$ and $\tauz\tauz$, all of which are even operators associated with the energy $\epsilon_{CFT}$ operator. This symmetry restriction can provide valuable clues despite the error bars on $\sigma$ and the decomposition coefficients being large. Finally, we also note that there are a lot of subtleties involved in the interpretation of the coefficients in the decomposition and their error bars (see Appendix~\ref{appendix:opdecompcoeffs} for a discussion).

\begin{table}[]
    \centering
    \begin{tabular}{|c|c|}
        \hline
        \textbf{Lattice Operator} & \textbf{Operator Decomposition} \\ \hline
        $\tauz$     & $-0.803121\sigma + 0.017\partial^2_{\tau}\sigma + 0.033\partial^2_x\sigma$ \\ \hline
        $\tauzxxz$ & $-0.803121\sigma + 0.820\partial^2_{\tau}\sigma + 0.736\partial^2_x\sigma$ \\ \hline
        $\tauy$    & $0.8031i\partial_{\tau}\sigma$ \\ \hline
        $\taux$     & $0.636620\mathbb{1} - 0.15915(T + \overline{T})$ \\
                & $- 0.31831\epsilon + 0.010\partial^2_{\tau}\epsilon$ \\ \hline
        $\tauz\tauz$     & $0.636620\mathbb{1} - 0.15915(T + \overline{T})$ \\
                & $+ 0.31831\epsilon - 0.010\partial^2_{\tau}\epsilon$ \\ \hline
        $\tauy\tauy$     & $-0.212207\mathbb{1} - 0.4774(T + \overline{T})$ \\
                & $+ 0.31831\epsilon - 0.089\partial^2_{\tau}\epsilon$ \\ \hline
        $\taux\taux$    & $0.540380\mathbb{1} - 0.5403(T + \overline{T})$ \\
                & $- 0.54038\epsilon + 0.067\partial^2_{\tau}\epsilon - 0.051\partial^2_{x}\epsilon$ \\ \hline
        $\taux\tauy+\tauy\taux$ & $2.41i\partial_\tau \sigma$ \\ \hline
    \end{tabular}
    \caption{1D TFIM: Expansions for lattice operators in terms of CFT operators, reproduced from~\cite{zouConformalFieldsOperator2020}}
    \label{tab:1DTFIM_OPE}
\end{table}

\subsection{2D Transverse Field Ising Model}
\label{sec:2dTFIM}

We now consider the two-dimensional (2D) Transverse Field Ising Model (2D TFIM):

\begin{equation}\label{eq:2d_TFIM}
    H = -J \sum_{\braket{ij}} \tauz_i \tauz_j - h \sum_{i} \taux_i,
\end{equation}
i.e. the same Hamiltonian as \eqref{eq:1DTFIM_Hamiltonian} except that spins are located on the 2D square lattice with the summation $\braket{ij}$ over the nearest-neighbor bonds. $J=1$ is the unit of energy and periodic boundary conditions are assumed. This model has also been extensively studied~\cite{Dutta_Aeppli_Chakrabarti_Divakaran_Rosenbaum_Sen_2015} and although there is no exact solution, numerical data indicate a quantum phase transition for $h_c=3.044330(6)$ in the $(2+1)$-d Ising universality class. In particular, the critical exponents are known with very high accuracy: $\nu =0.6299709750(12)$, $\eta = 0.03629761200(48)$\cite{changBootstrapping3dIsing2025, simmons-duffinLightconeBootstrapSpectrum2017, huangWormalgorithmtypeSimulationQuantum2020}.

\begin{figure}
    \centering
    \includegraphics[width=\linewidth]{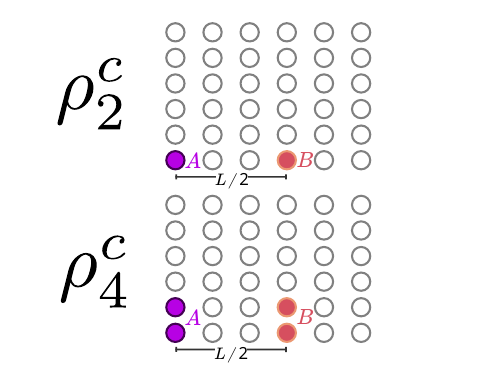}
    \caption{2D TFIM: Layout of clusters $A$ and $B$ for 4-site and 2-site CDMs.}
    \label{fig:2dTFIM_Layout}
\end{figure}

\begin{figure}
    \centering
    \includegraphics[width=\linewidth]{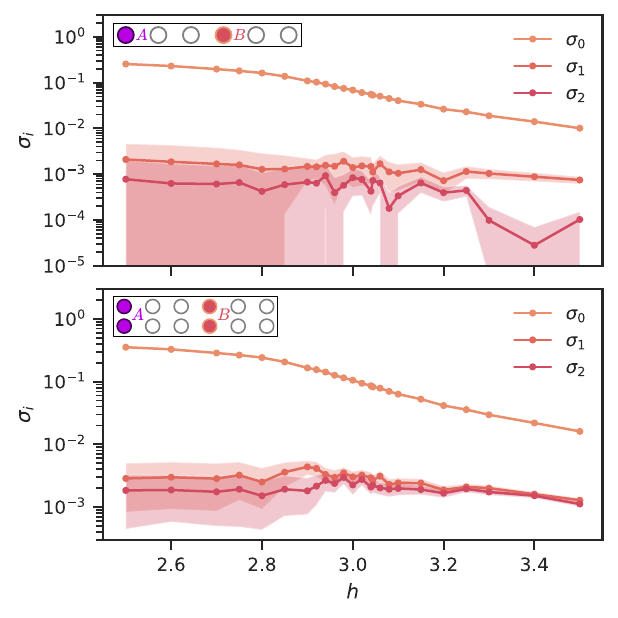}
    \caption{2D TFIM: Singular values vs field for 2-site (top) and 4-site (bottom) CDM for $L=8$}
    \label{fig:2dTFIM_SVs_vs_h}
\end{figure}

\begin{figure}[]
    \centering
    \begin{subfigure}[b]{0.8\linewidth}
        \centering
        \includegraphics[width=\linewidth]{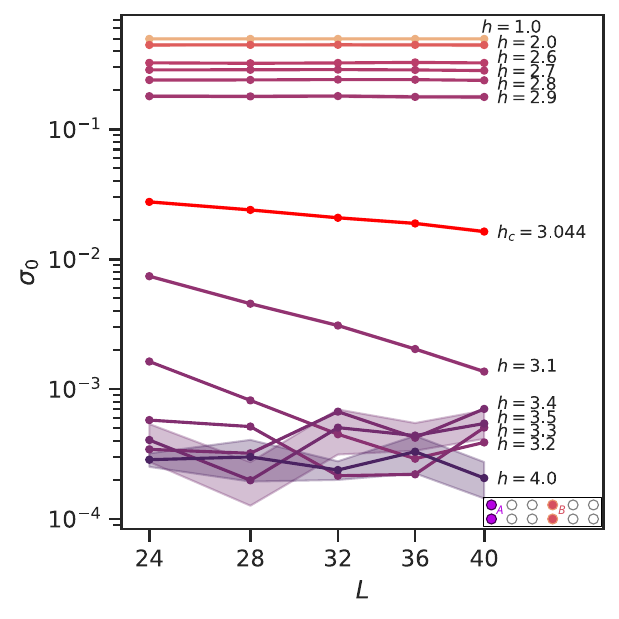}
        \label{fig:2dTFIM_near_hc}
        \caption{}
    \end{subfigure}
    \hfill
    \begin{subfigure}[b]{0.8\linewidth}
        \centering
        \includegraphics[width=\linewidth]{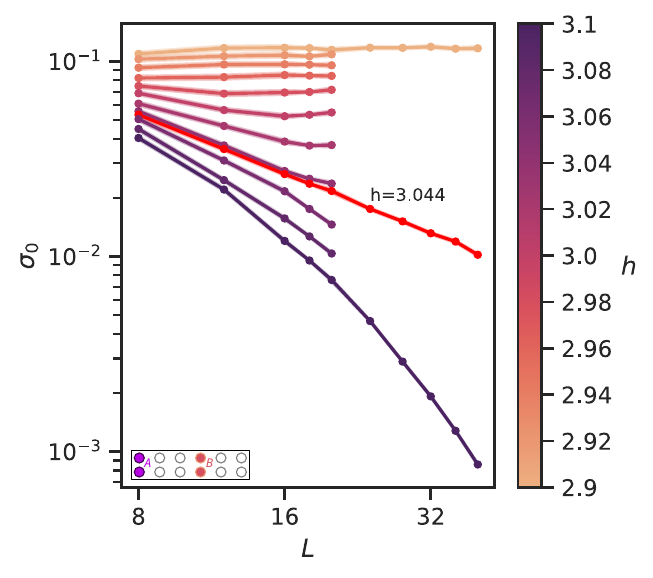}
        \label{fig:2dTFIM_near_hc_fine}
        \caption{}
     \end{subfigure}
    \caption{2D TFIM: Log-Log plot of $\sigma_0$ obtained from 4-site CDM vs $L$ for different values of field $h$ showing a separatrix indicative of the phase transition. Line colours indicate different values of field $h$. The lower panel has a narrower field range around the critical one at $h_c\simeq 3.044$.}
    \label{fig:2dTFIM_SV_vs_L_near_hc}
\end{figure}

We proceed in a similar fashion to the 1D case using a $\tauz$ computational basis. In this example, we choose site clusters which are non-collinear and exhibit multiple reflection symmetries. Considering two sets of sites $A$ and $B$ at opposite ends as shown in Fig.~\ref{fig:2dTFIM_Layout}, we use the SSE method to calculate the RDM and from it, the CDM. Decomposing this CDM using Eq.~\eqref{eq:cdmdecomp}, we plot the highest singular values for $\rho^c_4$ and $\rho^c_2$ (see Fig.~\ref{fig:2dTFIM_SVs_vs_h}). The singular values indicate that there are two phases for low and high values of the field $h$. The phase transition is indicated by the rapid decrease of $\sigma_0$ separating the ``ordered" phase and ``disordered" phase. This is verified by looking at the singular values as a function of field strength $h$ and system size $L$ as shown in Fig.~\ref{fig:2dTFIM_SV_vs_L_near_hc}. The $\sigma_0$ values stay nearly constant in the ``ordered" phase while they decay to the noise floor in the ``disordered" phase. The separatrix is found for $h_c\simeq 3.044$ in perfect agreement with the known value. 

The decomposition of the  operator $X_0$ corresponding to $\sigma_0$ shows that the system is dominated by $\braket{\tauz\tauz}_c$ correlations in the ordered phase. For the 4-site case, the dominant operator is a linear combination of $\tauzxxz$ and $\tauz \mathbb{1} + \mathbb{1} \tauz$ similar to the 1D TFIM case. We can consider $\sigma_0$ to be representative of an order parameter in this phase. In contrast, all singular values are very small and barely distinct beyond the phase transition. This implies the absence of any kind of ordering in this region. Most singular values become ill-defined in this region due to the intrinsic noise in our stochastic sampling.

The critical point exhibits a more intricate structure than the phases. The singular values of the CDM show a power law scaling in accordance with known CFT results. We argue that the operator decompositions of the different $X_i$ correspond to different leading scaling operators from the CFT if the corresponding $\sigma_i$ is well-defined. So each operator in $X_1$ for instance, when expressed in terms of the CFT operators and their derivatives, have the same leading order operator term which follows the same symmetries as the constituent operators in $X_1$ if $\sigma_1$ is well-defined.

\begin{figure}[]
    \centering
    \begin{subfigure}[b]{0.49\linewidth}
        \centering
        \includegraphics[width=\textwidth]{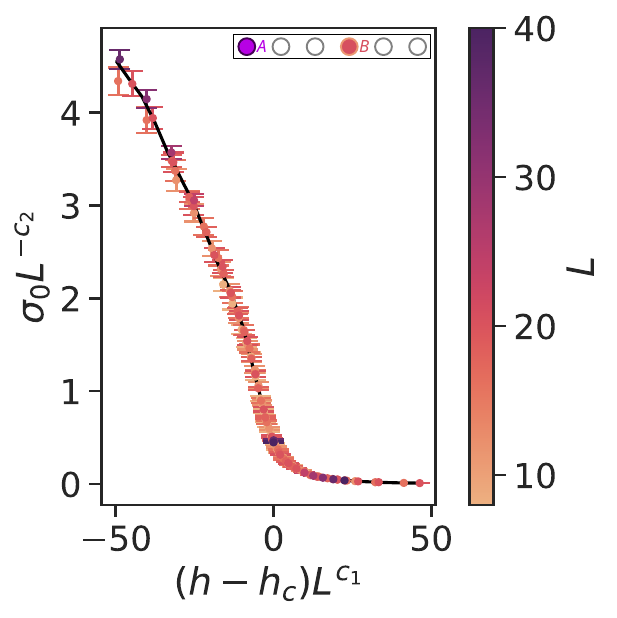}
        \caption{}
        \label{fig:2dTFIM_ScalingHarada}
    \end{subfigure}
    \hfill
    \begin{subfigure}[b]{0.49\linewidth}
        \centering
        \includegraphics[width=\textwidth]{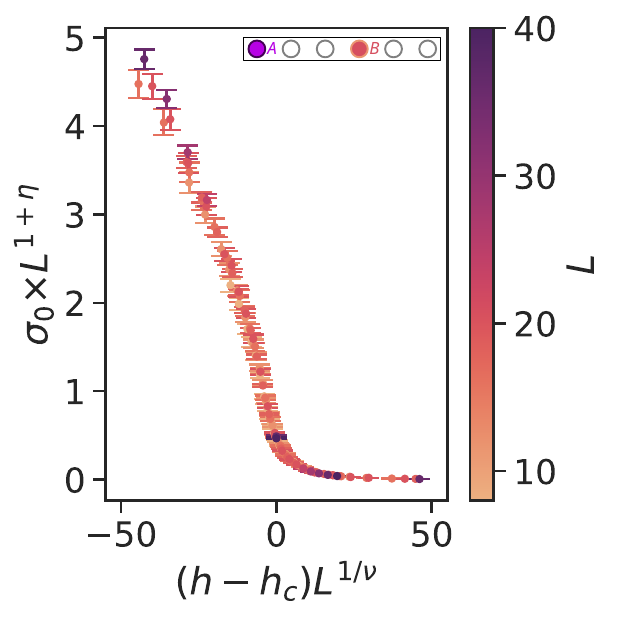}
        \caption{}
        \label{fig:2dTFIM_ScalingKnown}
    \end{subfigure}
    \caption{Data collapse for the 2D TFIM: (left) Finite size scaling fits $h_c^* = 3.04396(41)$, $\eta^* = 0.025(11)$, $\nu^* = 0.6153(23)$, (right) Using known values of critical point $h_c=3.044330(6)$ and scaling exponents $\nu =0.6299709750(12)$, $\eta = 0.03629761200(48)$~\cite{changBootstrapping3dIsing2025, simmons-duffinLightconeBootstrapSpectrum2017, huangWormalgorithmtypeSimulationQuantum2020}}
    \label{fig:2dTFIM_StdScaling}
\end{figure}

\begin{figure}[]
    \centering
    \includegraphics[width=\linewidth]{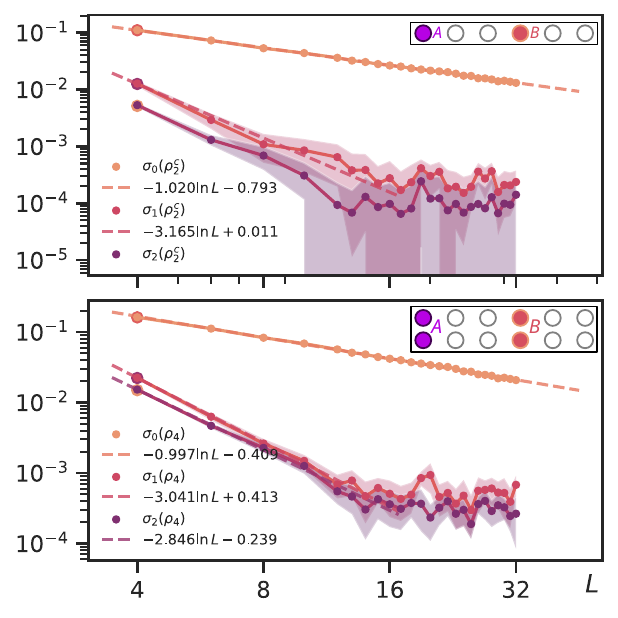}
    \caption{2D TFIM: 2-site and 4-site singular values at critical point. Smaller circles indicate values from the SSE at $\beta=2L$. Larger circles indicate values from exact diagonalization at $\beta=2L$ for the corresponding singular values.}
    \label{fig:2dTFIM_SV_vs_L}
\end{figure}

We expect to see the singular values (and correlation functions) decay with $L^{-2\Delta}$ where $\Delta$ is the corresponding CFT scaling dimension. The dominant one is known to be $2 \Delta_{\sigma} = 1.036297612(48)$ followed by $2 \Delta_{\epsilon} = 2.82525056(58)$ etc.\cite{changBootstrapping3dIsing2025, simmons-duffinLightconeBootstrapSpectrum2017}. We can obtain these via non-linear fitting as shown in Fig.~\ref{fig:2dTFIM_StdScaling} or via linearly fitting power law decays to the singular values at the critical point, if known and ascertained, as shown in Fig.~\ref{fig:2dTFIM_SV_vs_L}.

From the non-linear fitting with the scaling ansatz in Fig.~\ref{fig:2dTFIM_StdScaling}, we obtain a reasonable fit for the highest singular value $\sigma_0$. The estimates of $h_c$ and $\eta$ are in quite good agreement with the best known values. Since $d + z=3$, the expected value of $c_2$ is $-(1 + \eta)$ and the expected value of $c_1$ is $1/\nu$. After locating the critical point, we can obtain a better fit for the critical exponents by considering the decay of the connected correlation of the order parameter, in this case $\sigma_0$, at the critical point as a function of system size.

Quite remarkably, as shown in Fig.~\ref{fig:2dTFIM_SV_vs_L}, our numerical data on the dominant singular value leads to a power-law exponent close to $-1.020(8)$ in very good agreement with $2\Delta_\sigma$, while the subleading data (with more noise) show a power-law decay with an exponent close to $-2.85(6)$, consistent with the known value of $2\Delta_\epsilon$. For subleading terms, the fitting is done for smaller lengths and hence show finite size effects - correlations with larger decay exponents may be higher in magnitude than those with smaller decay exponents.

\begin{table}[]
    \begin{tabular}{|c|cc|cc|cc|}
        \hline
        \textbf{Op} & \multicolumn{2}{l|}{\textbf{Ordered}  $\mathbf{h = 1}$} & \multicolumn{2}{l|}{\textbf{Critical} $\mathbf{h = 3.044}$} & \multicolumn{2}{l|}{\textbf{Disordered} $\mathbf{h = 3.5}$} \\ \hline
%        Op &
%          \multicolumn{2}{c|}{$h= 1$} &
%          \multicolumn{2}{c|}{$h=3.044$} &
%          \multicolumn{2}{c|}{$h=3.5$} \\ \hline
        $X_0$ &
          \multicolumn{1}{c|}{$1.0^\# \odd{ \frac{\tauz}{\sqrt{2}} }$} &
          $0.468(7)$ &
          \multicolumn{1}{c|}{$1.0^\# \odd{ \frac{\tauz}{\sqrt{2}} }$} &
          $0.0533(14)$ &
          \multicolumn{1}{c|}{$1.0^\# \odd{ \frac{\tauz}{\sqrt{2}} }$} &
          $0.01010(19)$ \\ \hline
        $X_1$ &
          \multicolumn{1}{c|}{$1.0i^* \odd{ \frac{\tauy}{\sqrt{2}} }$} &
          $0.002(2)$ &
          \multicolumn{1}{c|}{$-1.0i^* \odd{ \frac{\tauy}{\sqrt{2}} }$} &
          $0.0011(6)$ &
          \multicolumn{1}{c|}{$-1.0i^\# \odd{ \frac{\tauy}{\sqrt{2}} }$} &
          $0.0007(1)$ \\ \hline
        $X_2$ &
          \multicolumn{1}{c|}{$1.0^* \even{ \frac{\taux}{\sqrt{2}} }$} &
          $0.001(2)$ &
          \multicolumn{1}{c|}{$1.0^* \even{ \frac{\taux}{\sqrt{2}} }$} &
          $0.0007(5)$ &
          \multicolumn{1}{c|}{$-1.0^\# \even{ \frac{\taux}{\sqrt{2}} }$} &
          $0.00010(5)$ \\ \hline
    \end{tabular}
    \caption{2D TFIM: Operator Decomposition from 2-site CDM at $L=8$. $*$ denotes that the error is of the order of the value. $\#$ denotes that there is no error within the bootstrap samples.}
    \label{tab:2dTFIM_OpDecomp_2Site}
\end{table}

% \onecolumngrid

\begin{table*}
    \centering
    \begin{tabular}{|c|ll|ll|ll|}
        \hline
        \textbf{Op} & \multicolumn{2}{l|}{\textbf{Ordered}  $\mathbf{h = 1}$} & \multicolumn{2}{l|}{\textbf{Critical} $\mathbf{h = 3.044}$} & \multicolumn{2}{l|}{\textbf{Disordered} $\mathbf{h = 3.5}$} \\ \hline
        $X_0$ &
          \multicolumn{1}{p{0.21\textwidth}|}{$0.6863(2) \odd{(\mathbb{1}\tau^z + \tau^z\mathbb{1})/2} + 0.1704(9) \odd{(\tau^x\tau^z + \tau^z\tau^x)/2}$} &
          0.497(4) &
          \multicolumn{1}{p{0.21\textwidth}|}{$0.5644(4)\odd{(\tau^z\mathbb{1} + \mathbb{1}\tau^z)/2} + 0.4260(6) \odd{(\tau^z\tau^x + \tau^x\tau^z)/2}$} &
          % 0.0824(15) &
          0.082(2) &
          \multicolumn{1}{p{0.21\textwidth}|}{$0.5440(2)\odd{(\tau^z\mathbb{1} + \mathbb{1}\tau^z)/2} + 0.4517(3) \odd{(\tau^z\tau^x + \tau^x\tau^z)/2}$} &
          % 0.01608(22) \\ \hline
          0.0161(2) \\ \hline
        $X_1$ &
          \multicolumn{1}{p{0.21\textwidth}|}{$-0.175i^* \odd{(\tau^x\tau^y + \tau^y\tau^x)/2}
          - 0.685i^* \odd{(\mathbb{1}\tau^y + \tau^y\mathbb{1})/2}$} &
          % 0.0024(17) &
          0.002(2) &
          \multicolumn{1}{p{0.21\textwidth}|}{$-0.5^* \even{\taux\taux/2} + 0.110^* \even{\tauy\tauy/2} + 0.574^* \even{\tauz\tauz/2}  - 0.392^* \even{(\mathbb{1}\taux + \taux\mathbb{1})/2}$} &
          0.0027(5) &
          \multicolumn{1}{p{0.21\textwidth}|}{$0.51(8)\even{(\tauz\taux -\taux\tauz)/2} + 0.48(7)\even{(\tauz\mathbb{1} - \mathbb{1}\tauz)/2}$} &
          % 0.00128(10) \\ \hline
          0.0013(1) \\ \hline
        $X_2$ &
          \multicolumn{1}{p{0.21\textwidth}|}{$0.696^* \odd{(\tau^z\tau^x - \tau^x\tau^z)/2} + 0.127^* \odd{(\tau^z\mathbb{1}/2 - \mathbb{1}\tau^z)/2}$} &
          0.0015(9) &
          \multicolumn{1}{p{0.21\textwidth}|}{$-0.548i^* \odd{(\tau^x\tau^y + \tau^y\tau^x)/2} - 0.447i^* \odd{(\mathbb{1}\tau^y + \tau^y\mathbb{1})/2}$} &
          0.0022(5) &
          \multicolumn{1}{p{0.21\textwidth}|}{$-0.49(4)i \odd{(\mathbb{1}\tauy + \tauy\mathbb{1})/2} - 0.51(5)i\odd{(\taux\tauy + \tauy\taux)/2}$} &
          % 0.00112(10) \\ \hline
          0.0011(1) \\ \hline
    \end{tabular}
    \caption{2D TFIM: Operator decomposition for 4-site CDM at $L=8$ with corresponding singular values. $*$ denotes that the error is of the order of the value.}
    \label{tab:2dTFIM_OpDecomp_4Site}
\end{table*}

% \twocolumngrid

\subsection{Heisenberg Bilayer}\label{sec:bilayer}
We now consider the bilayer spin-1/2 Heisenberg antiferromagnetic (AF) model 
\begin{align}
    H &= J_h \sum_{ \substack{\braket{ij} \\ ij \in P}} \mathbf{S_i} \cdot \mathbf{S_j} 
    + J_v \sum_{ \substack{ \braket{ij} \\ i \in P_1 \\ j \in P_2} } \mathbf{S_i} \cdot \mathbf{S_j}
\end{align}
where $J_h$ and $J_v$ are the AF nearest-neigbor couplings within/between each plane respectively. We consider square lattice geometry for each plane $P_1$ and $P_2$ with periodic boundary conditions and even linear size $L$. We fix $J_h = 1$ and vary the interlayer-intralayer coupling ratio $g = J_v/J_h = J_v$. This model is known to exhibit an O(3) quantum phase transition in its ground state as a function of $g$, in the same universality class as the finite temperature three-dimensional (3D) classical Heisenberg model. For small $g$ the system shows long-range N\'eel ordering, this is destroyed with increasing $g$ by the transition through the opening of a spin gap. For large $g$, the ground state tends towards a product state of rung singlets between the two layers which is a trivial state (i.e. unique, with a finite gap and hence no long-range correlations). The critical value $g_c=2.5220(1)$ is known very accurately from earlier QMC simulations~\cite{wangHighprecisionFinitesizeScaling2006}.

\begin{figure}
    \centering
    \includegraphics[width=\linewidth]{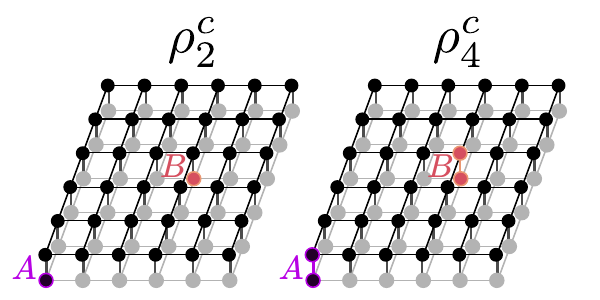}
    \caption{Heisenberg bilayer: Layout of clusters $A$ and $B$ for 2-site and 4-site CDM.}
    \label{fig:HeisBilayer_Layout}
\end{figure}

Since the model admits a global SU(2) rotation symmetry, we choose a basis of SU(2) invariant quantities for decomposing the 4-site and 2-site CDMs. The construction of this basis can be found in Ref~\cite{sudanUncoveringPhysicsFrustrated2010}. For the geometry, we choose 2 rungs on diagonally opposite ends of the plane for maximal distance and to characterize the long-range dimer-dimer pattern at large $g$ (see Figure \ref{fig:HeisBilayer_Layout}). We impose total $S_z$ conservation and geometrical reflection symmetries during the calculation of the error bars.

\begin{figure}
    \centering
    \includegraphics[width=0.95\linewidth]{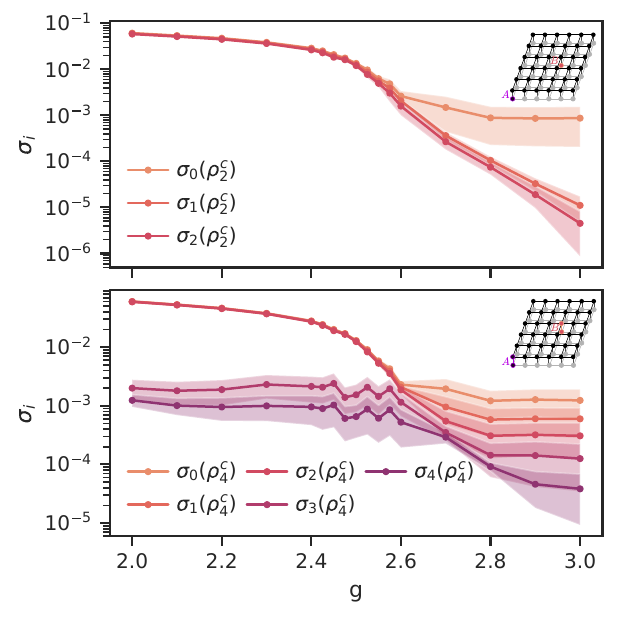}
    \caption{Heisenberg bilayer: Singular values from 2-site and 4-site CDM vs coupling ratio $g$ for $L=32$.}
    \label{fig:HeisBilayer_SV_vs_g}
\end{figure}

\begin{figure}
    \centering
    \includegraphics[width=0.95\linewidth]{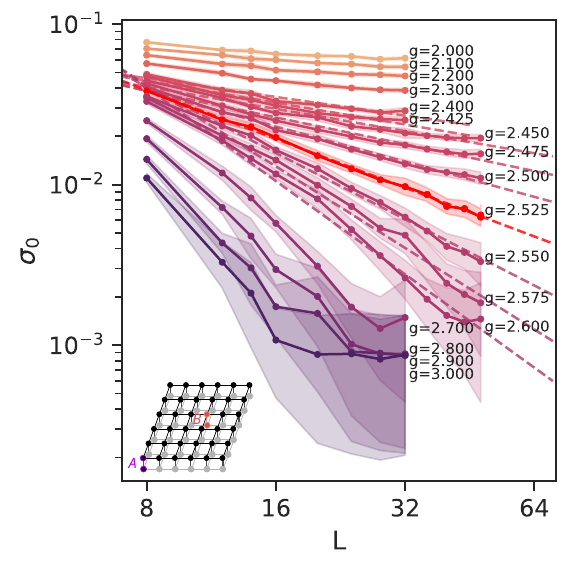}
    \caption{Heisenberg bilayer: Highest singular value from 4-site CDM vs linear size for different values of coupling ratio}
    \label{fig:HeisBilayer_SV_vs_L}
\end{figure}

From the SVD, we see that the three-fold degenerate dominant correlation $\sigma_0$ corresponds to the triplet operator $(\mathbf{S}-\mathbf{S})$ for $g < g_c$, in agreement with the presence of a staggered N\'eel phase, whereas all correlations decay rapidly beyond $g_c$ such that we cannot discern a dominant correlation at our error resolution as shown in Fig.~\ref{fig:HeisBilayer_SV_vs_g}. This is exemplified in Fig.~\ref{fig:HeisBilayer_SV_vs_L} where the dominant correlations remain relatively unchanged in the ``ordered" phase and decay rapidly in the ``disordered" phase, the two of which are divided by the separatrix indicating the critical point. From this plot, our data lead to a critical value $g_c=2.525(25)$ in excellent agreemeent with earlier studies~\cite{wangHighprecisionFinitesizeScaling2006}.
Therefore we classify the two phases as ordered and disordered respectively. We note that our singular value spectrum reflects the SU(2) symmetry through its degeneracies - for the 4-site CDM, we expect to see 2 non-degenerate, 3 sets of triply degenerate and 1 set of quintuply degenerate singular values - this can be observed in the spectrum within the error resolution. We see that the three highest singular values are degenerate and the mean CDM decomposition shows us that these operators correspond to a mix of $(\mathbf{S} - \mathbf{S})$ components, as expected for a triplet operator. Since $\sigma_{3+}$ are ill-defined, it is hard to comment on their nature directly.  

\begin{figure}[]
    \centering
    \begin{subfigure}[b]{0.49\linewidth}
        \centering
        \includegraphics[width=\textwidth]{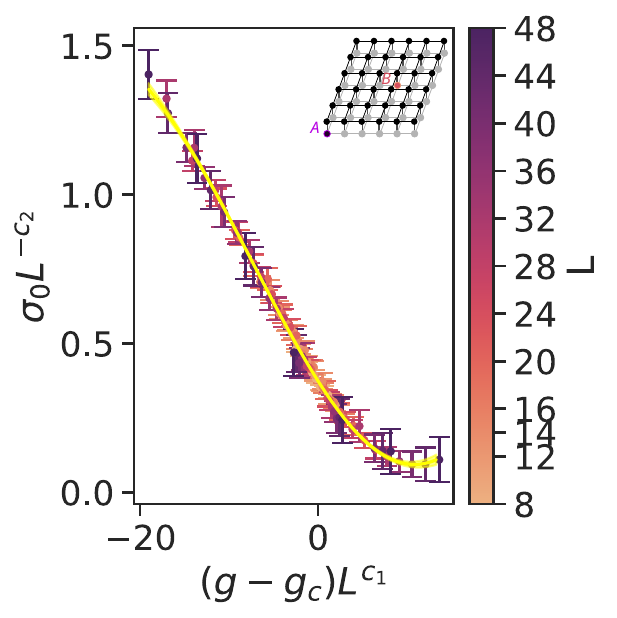}
        \caption{}
        \label{fig:HeisBilayer_ScalingHarada}
    \end{subfigure}
    \hfill
    \begin{subfigure}[b]{0.49\linewidth}
        \centering
        \includegraphics[width=\textwidth]{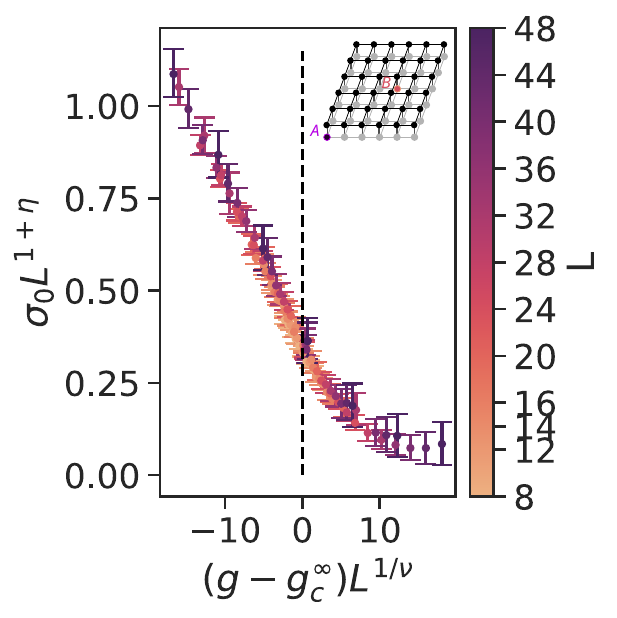}
        \caption{}
        \label{fig:HeisBilayer_ScalingKnown}
    \end{subfigure}
    % \caption{Data collapse for the Heisenberg Bilayer: (left) Finite Size Scaling Fits $g_c = 2.5195(52)$, $\eta = -0.061(33)$, $\nu = 0.680(17)$, (right) Using known values of critical point and scaling exponents $g_c = 2.5220(1)$, $\nu = 0.7112(5)$, $\eta = 0.0375(5)$. \cite{campostriniCriticalExponentsEquation2002, wangHighprecisionFinitesizeScaling2006}}
    \caption{Data collapse for the Heisenberg bilayer: (a) Finite size scaling leads to $g_c^* = 2.537(5)$, $\eta^* = 0.10(4)$, $\nu^* = 0.72(3)$; (b) Using known values of critical point and scaling exponents $g_c = 2.5220(1)$, $\nu = 0.7112(5)$, $\eta = 0.0375(5)$. \cite{campostriniCriticalExponentsEquation2002, wangHighprecisionFinitesizeScaling2006}}
    \label{fig:HeisBilayer_Scaling}
\end{figure}

Using the connected correlation of $X_0$ as the order parameter, the finite size scaling of the dominant correlation $\sigma_0$ yields a reasonable fit for the critical exponents and parameters as shown in Fig.~\ref{fig:HeisBilayer_ScalingHarada}. With the non-linear fitting of the scaling ansatz, we obtain a reasonable estimate for the critical point $g_c$ and critical exponents $\nu$ and $\eta$. The known values for these critical exponents should match with $c_2 = -(1 + \eta)$ and $c_1 = 1/\nu$, and this comparison indicates a reasonable agreement.

% \onecolumngrid

\begin{table*}[]
    \begin{tabular}{|l|p{0.2\linewidth}p{0.1\linewidth}|p{0.2\linewidth}p{0.1\linewidth}|p{0.2\linewidth}p{0.1\linewidth}|}
        \hline
        \textbf{Op} & \multicolumn{2}{l|}{ \textbf{Ordered} $\mathbf{g=2}$ } & \multicolumn{2}{l|}{\textbf{Critical} $\mathbf{g=2.525}$} & \multicolumn{2}{l|}{\textbf{Disordered} $\mathbf{g=3}$} \\ \hline
        $X_0$ &
          \multicolumn{1}{l|}{$1.0 \sqrt{2}S_z$} &
          $0.061(1)$ &
          \multicolumn{1}{l|}{$-0.7071 \sqrt{2}(S_x + i S_y)$} &
          $0.010(1)$ &
          \multicolumn{1}{l|}{$0.7071 \sqrt{2}(\mathbb{1}/2 - S_z)$} &
          $9(7) \times 10^{-4}$ \\ \hline
        $X_1$ &
          \multicolumn{1}{l|}{$-0.7101 \sqrt{2}(S_x + i S_y)$} &
          $0.060(1)$ &
          \multicolumn{1}{l|}{$0.7071 \sqrt{2}(i S_y - S_x)$} &
          $0.0087(4)$ &
          \multicolumn{1}{l|}{$0.7071 \sqrt{2}(S_x - i S_y)$} &
          $1.1(6) \times 10^{-5}$ \\ \hline
        $X_2$ &
          \multicolumn{1}{l|}{$0.7101 \sqrt{2}(i S_y - S_x)$} &
          $0.058(1)$ &
          \multicolumn{1}{l|}{$1.0 \sqrt{2} S_z$} &
          $0.008(1)$ &
          \multicolumn{1}{l|}{$0.7071 \sqrt{2}(S_x + i S_y)$} &
          $5(4) \times 10^{-6}$ \\ \hline
    \end{tabular}
    \caption{Heisenberg bilayer: Operator decomposition and corresponding singular value from 2-site CDM at $L=32$}
    \label{tab:HeisBilayer_OpDecomp_2Site}
\end{table*}

\begin{table*}[]
    \centering
    \begin{tabular}{|l|p{0.2\linewidth}p{0.1\linewidth}|p{0.2\linewidth}p{0.1\linewidth}|p{0.2\linewidth}p{0.1\linewidth}|}
        \hline
        \textbf{Op} &
          \multicolumn{2}{l|}{\textbf{Ordered} $\mathbf{g=2}$} &
          \multicolumn{2}{l|}{\textbf{Critical} $\mathbf{g=2.525}$} &
          \multicolumn{2}{l|}{\textbf{Disordered} $\mathbf{g=3}$} \\ \hline
        $X_0$ &
          \multicolumn{1}{p{0.2\linewidth}|}{$1.0 (S-S)_z$} &
          $0.0604(6)$ &
          \multicolumn{1}{p{0.2\linewidth}|}{$0.7097 (S-S)_x + 0.7045i(S-S)_y$} &
          $0.0091(5)$ &
          \multicolumn{1}{p{0.2\linewidth}|}{$-0.0003332 Q_2 - 1.0 (S \cdot S)$} &
          $0.0012(7)$ \\ \hline
        $X_1$ &
          \multicolumn{1}{p{0.2\linewidth}|}{$-0.7101 ((S-S)_x + i(S-S)_y)$} &
          $0.0600(3)$ &
          \multicolumn{1}{p{0.2\linewidth}|}{$0.7045 (S-S)_x - 0.7097i(S-S)_y$} &
          $0.00874(7)$ &
          \multicolumn{1}{p{0.2\linewidth}|}{$3.332 \times^{-4} (S \cdot S) - 1.0 Q_2$} &
          $0.0006(3)$ \\ \hline
        $X_2$ &
          \multicolumn{1}{p{0.2\linewidth}|}{$0.7101 (i(S-S)_y - 0.7101 (S-S)_x)$} &
          $0.0595(7)$ &
          \multicolumn{1}{p{0.2\linewidth}|}{$1.0 (S-S)_z$} &
          $0.0083(6)$ &
          \multicolumn{1}{p{0.2\linewidth}|}{-} &
          $0.0003(2)$ \\ \hline
        $X_3$ &
          \multicolumn{1}{p{0.2\linewidth}|}{$0.831 Q_3 - 0.556 Q_1$} &
          $0.0020(8)$ &
          \multicolumn{1}{p{0.2\linewidth}|}{$9.716 \times 10^{-5} Q_2 - 1.0 (S \cdot S)$} &
          $0.002(1)$ &
          \multicolumn{1}{p{0.2\linewidth}|}{-} &
          $0.00013(9)$ \\ \hline
        $X_4$ &
          \multicolumn{1}{p{0.2\linewidth}|}{$0.831 Q_1 + 0.556 Q_3$} &
          $0.0012(3)$ &
          \multicolumn{1}{p{0.2\linewidth}|}{$-1.0 (S \times S)_z$} &
          $0.0009(6)$ &
          \multicolumn{1}{p{0.2\linewidth}|}{-} &
          $4(3) \times 10^{-5}$ \\ \hline
    \end{tabular}
    \caption{Heisenberg bilayer: Operator decomposition and corresponding singular value for 4-site CDM at $L=32$}
    \label{tab:HeisBilayer_OpDecomp_4Site}
\end{table*} 

% \twocolumngrid

%%%%%%%%%%%%%%%%%%%%%%%%%%%%%%%%%%%%%%%%%%%%%%%%%%%%%%%%%%%%%%%
\section{Conclusion}\label{sec:conclusion}
We have introduced a quantum Monte Carlo-based approach to find the relevant dominant correlations in any given phase of an interacting quantum spin system in any dimension. By utilizing the correlation density matrix-based tools, we have opened up the possibility of discovering new, exotic order parameters in novel phases of matter. The approach allows us to characterize any phase of matter amenable to SSE by its dominant connected correlations in an unbiased fashion. We have introduced reliable methods of understanding the uncertainties in the quantities resulting from this stochastic approach. 

Furthermore, this approach can be used to detect quantum phase transitions and perform finite size scaling analysis, as shown on some paradigmatic models (1D and 2D transverse field Ising model, bilayer Heisenberg antiferromagnet), with very good accuracy. This has enabled us to locate the phase transitions and determine the associated critical exponents. Using this approach, we have also obtained information about the leading terms in the CFT operator expansions of the lattice operators at quantum critical points. Finally, the approach introduced allows for post-simulation access to all the connected correlations between the chosen clusters $A$ and $B$.

While this approach is limited by the sizes of $A$ and $B$ in computation time, it offers immense flexibility in terms of the geometry of these clusters. Future work can extend the number of sites by exploiting geometrical symmetries. It invites further studies on the entanglement-related measures such as multi-partite entanglement and topological entanglement entropy. The approach also has the potential to explore possibilities between dominant ordering schemes on different length scales by changing the distance between $A$ and $B$ since the distance separating $A$ and $B$ does not affect the simulation in any significant manner. While this approach primarily targets detection and characterization of long-range order, it also enables the identification of the lack of correlations, which is characteristic of quantum spin liquids~\cite{cheongCorrelationDensityMatrix2009}, provided they are amenable to SSE. Several QMC-accessible models are known to host quantum spin liquids~\cite{balentsFractionalizationEasyaxisKagome2002,wangQuantumSpinLiquid2018,hermelePyrochlorePhotons$U1$2004,banerjeeUnusualLiquidState2008,blockKagomeModel$mathbbZ_2$2020}. Since gapped quantum spin liquids exhibit topological order, future work could employ geometries that encompass the boundary of a non-local region, similar to the construction for topological entanglement entropy~\cite{levinDetectingTopologicalOrder2006,kitaevTopologicalEntanglementEntropy2006}. A CDM defined on such a geometry may capture topological information and thus differentiate quantum spin liquids from a disordered state. The associated RDM would also yield the topological entanglement entropy. Although this requires subsystem size scaling as $|AB| \sim O(L)$, the computational cost may be alleviated by exploiting translation symmetries, as demonstrated in~\cite{maoSamplingReducedDensity2025}. 

In the future, it would be useful to see if one can use similar ideas to detect topological phases, which by essence do not have local order parameter, e.g. by using nonlocal measurements such as the string order parameter~\cite{NijsRommelse1989} in the spin-1 Haldane phase or considering 1d or 2d symmetry-protected topological phases~\cite{Wen2017} or measuring hidden order in fractional Chern insulators~\cite{Pauw2024}. Quite interestingly, some famous quantum materials do possess some hidden order which has not been characterized by conventional experimental probes despite decades of research~\cite{URuSi}. 

{\it Acknowledgments -- } We acknowledge insightful discussions with Andreas L\"auchli, Zheng Yan and Jiarui Zhao. 
We thank CALMIP (grant 2025-P0677) and GENCI (project A0150500225) for computer resources.
We acknowledge support from the ANR/RGC Joint Research Scheme sponsored by Research Grants Council of Hong Kong SAR of China (Project No. AHKU703/22) and French National Research Agency (grant ANR-22-CE30-0042-01).

\bibliography{main}
% using biblatex
% \printbibliography

%%%%%%%%%%%%%%%%%%%%%%%%%%%%%%%%%%%%%%%%%%%%%%%%%%%%%%%%%%%%%%%
\appendix
\section{}
\subsection{Decomposition of the correlation density matrix}
\label{appendix:svd}

We reshape the correlation density matrix to separate the indices on $A$ and $B$:
\begin{align}
\left( \rho^c_{AB} \right)_{i,j} = \left( \rho^c_{AB} \right)_{\underline{ab},\underline{a'b'}} \rightarrow \tilde{\rho}_{\underline{aa'},\underline{bb'}}
\end{align}
Using a singular value decomposition, we can write:
\begin{align}
\tilde{\rho} &= U \Sigma V^\dagger \\
\left( X_i^{(A)} \right)_{a, a'} &= U_{\underline{aa'},i} \\
\left( Y_j^{(B)} \right)^\dagger_{b, b'} &= V^\dagger_{j, \underline{bb'}} \\
\sigma_i^{AB} &= \Sigma_{ii} 
\end{align}

The operators are orthonormal under the Frobenius norm: $\mathrm{Tr}[ X_i^{(A)} (X_j^{(A)})^\dagger ] = \mathrm{Tr}[ Y_i^{(B)} (Y_j^{(B)})^\dagger ] = \delta_{ij}$.

\subsection{Details of simulations and analysis}

\subsubsection{Simulation details}
\label{sec:SimulationDetails}

Since we are interested in the groundstate properties of the models considered, we seek to suppress Monte Carlo moves that correspond to energy changes greater than the finite-size energy gap. It is, therefore, important to choose the value of the inverse temperature $\beta$ appropriately and scale it with system size to ensure $e^{-\beta \Delta} << 1$ for phases with energy gap $\Delta$. In the two Ising models, we use the second energy gap in the ordered phase since the first excited state is quasi-degenerate with the ground state. For 1D Ising, we use $\beta = L$ with the 10,000,000 sweeps of the SSE algorithm. For 2D Ising and Heisenberg Bilayer models, we use $\beta = 2L$ and 10,000,000 sweeps of the SSE algorithm. For the bootstrap algorithm, we generate 1000 samples of the CDM matrix and draw 999 bootstrap sets to draw our conclusions.

\subsubsection{Sampling of the CDM and Bootstrapping}
\label{sec:Bootstrap}
From the quantum Monte Carlo simulation, we obtain the reduced density matrix $\rho_{AB}$ along with standard deviations for each element of the matrix $\Delta \rho_{AB}$. In order to perform the SVD analysis with appropriate statistics, we utilize bootstrap-based methods~\cite{youngEverythingYouWanted2015}. We generate a set of samples $\{\rho_{AB}^{(i)}\}$ which are drawn from the distribution $\rho_{AB} \pm \Delta \rho_{AB}$ in a symmetry conserving manner. The analysis is then performed on each sample producing the set of results $\{Q^{(i)}\}$ where $Q$ is any quantity calculated in the analysis such as singular values of the CDM, the physical correlations or the coefficients of the decompositions of $X_i^{(A)}$. The final value is then calculated from the mean of this distribution and then error bar is calculated as the 95\% confidence interval of the mean via bootstrap. It is important that the symmetries be maintained at each step of the analysis. In particular at the order-disorder critical point, asymmetric deviations from the mean will result in values with high variance (for quantities that are finite in the ordered phase and close to zero in the disordered phase). This also leads to asymmetric terms in the decomposition of $X_i$'s. In order to avoid this, we symmetrize the $\rho_{AB}$ obtained from the Monte Carlo simulation, and then group the symmetric elements of the matrix in order to sample each group only once i.e. if elements $a_{ij} = a_{kl}$ by some symmetry then the sampled elements $\tilde{a}_{ij} = \tilde{a}_{kl}$. 

\subsubsection{Subtleties of the operator decomposition coefficients}
\label{appendix:opdecompcoeffs}

The operator decomposition lets us understand the phase through the dominant correlations, but the analysis has a few subtleties, some of which being linked to the fact that we only measure the CDM \emph{statistically}. 

\emph{What is a coefficient error bar? ---} The error bar on the operator decomposition coefficients is generated by sampling a distribution of possible RDMs, then applying the transformation to CDM and the SVD procedure on all matrices in this distribution. The coefficient of basis operator (such as a Pauli string) is then calculated for each $X_i$ obtained from these matrices through the Frobenius product. One of the major reasons for a high error bar on these coefficients is instability in the order of the $X_i$'s. Consider a system with independent correlations $\braket{(\kappa K + \gamma G)(\kappa K + \gamma G)}_c = \sigma_{KG}$ and $\braket{(\mu M + \eta N)} = \sigma_{MN}$. If $\sigma_{KG}$ and $\sigma_{MN}$ have overlapping error bars, which can occur through degeneracies or ill-definedness, some RDM samples will have $\sigma_{KG} > \sigma_{MN}$ whereas others will have $\sigma_{KG} < \sigma_{MN}$. As a result, some have $X_i = \kappa K + \gamma G$, $X_{i+1} = \mu M + \eta N$ whereas the others will have the $X_i = \mu M + \eta N$, $X_{i+1} = \kappa K + \gamma G$. The measurement of the coefficient of $K$ in the first case is $\kappa$, while in the second case it is $0$. This leads to a Bernoulli-like distribution which shows a large variance. If this occurs due to $\sigma_{KG}$ and $\sigma_{MN}$ being ill-defined, then the key takeaway from this is that the operator groups $\{K, G\}$ and $\{M, N\}$ follow the same symmetry and are likely (but not guaranteed) to be corresponding to the CFT operators of the same symmetry and order. They represent the shortcomings of this method and would require either longer simulations or other methods.

\emph{Sign of the coefficient ---} Consider an independent correlation $\braket{O_A O_B}_c = C_O$. This would be reflected in the SVD as $\braket{X_i^{(A)} (Y_j^{(B)})^\dagger}_c = \sigma_i$ for some $i$. $C_O$ can be negative but the SVD procedure guarantees that $\sigma$ is always positive. As a result, the factor of $-1$ is included in either $X_i^{(A)}$ or $(Y_j^{(B)})^\dagger$ with equal chance. If $X_i = \kappa K + \gamma G$, measuring the coefficient $\kappa$ directly would therefore lead to a case where the mean is close to zero and a very large variance. Instead the absolute value $|\kappa|$ should be measured and averaged over the bootstrap distribution. Note also that the relative sign between $\kappa$ and $\gamma$ remains unchanged throughout and thus $\kappa / \gamma$ can be used as well.

\emph{Degeneracies ---} In case of degeneracies, we may have a case where the coefficients have large variances owing to the inherent instability of the ordering i.e. if $K$, $M$, $N$ are operators with degenerate, independent correlations, then their ordering in the SVD from the CDM distribution will a mixture of all 6 possibilities. As a result, the relevant quantity to measure would be the sum of the squares of the absolute values of the coefficients of $K$, $M$ and $N$. This is seen in the SU(2) invariant case of the bilayer Heisenberg model with the presence of three triplets and one quintuplet in the singular value spectrum. 

\emph{Stability with system size ---} The operator decomposition interpretations we have mentioned hold only when the decomposition remains relatively stable with increasing system size. For instance, if $X_i = \kappa K + \gamma G$, we argue that if $\braket{KK}_c \sim L^\alpha$ then $\braket{GG}_c \sim L^\alpha$ must hold too. If it scales slower than $\alpha$, the coefficient $\gamma$ of $G$ should vanish with increasing system size (and vice versa). Therefore, it is important to establish stability of the decomposition with system size.

\end{document}